\documentclass[useAMS,usenatbib]{mn2e}
\usepackage{graphicx}
\usepackage[figuresright]{rotating}
\usepackage{float}
\usepackage{subfigure}
\usepackage{color}

\newcommand{\ewr}{\mbox{$W_r(1548)$}}
\newcommand{\apg}{\:^{>}_{\sim}\:}
\newcommand{\apl}{\:^{<}_{\sim}\:}

\newcommand{\etal}{et al.}

\newcommand{\kms}{\mbox{km\ s${^{-1}}$}}
\newcommand{\lya}{\mbox{${\rm Ly}\alpha $}}

\newcommand{\civ}{\mbox{${\rm C\,IV}$}}
\newcommand{\mkap}{\mbox{$\langle\kappa\rangle$}}

\begin{document}


\title[Mining Low-redshift Circumgalactic Baryons]{Mining Circumgalactic Baryons in the Low-Redshift Universe}

\author[Liang $\&$ Chen]{Cameron J.\ Liang$^{1}$\thanks{E-mail:
jwliang@oddjob.uchicago.edu}, Hsiao-Wen Chen$^{1}$\thanks{E-mail:
hchen@oddjob.uchicago.edu}\\
\\
$^{1}$Department of Astronomy \& Astrophysics, and Kavli Institute for Cosmological Physics, University of Chicago, Chicago IL 60637 }

\pagerange{\pageref{firstpage}--\pageref{lastpage}} \pubyear{2011}

\maketitle

\label{firstpage}

\begin{abstract}

  This paper presents an absorption-line study of the multiphase
  circumgalactic medium (CGM) based on observations of \lya, C\,II,
  C\,IV, Si\,II, Si\,III, and Si\,IV absorption transitions in the
  vicinities of 195 galaxies at redshift $z<0.176$.  The galaxy sample
  is established based on a cross-comparison between public galaxy and
  QSO survey data and is characterized by a median redshift of
  $\langle\,z\,\rangle = 0.041 $, a median projected distance of
  $\langle\,d\,\rangle=362$ kpc to the sightline of the background
  QSO, and a median stellar mass of $\log\,(M_{\rm star}/M_\odot) =
  9.7 \pm 1.1$.  Comparing the absorber features identified in the QSO
  spectra with known galaxy properties has led to strong constraints
  for the CGM absorption properties at $z\apl 0.176$.  First, abundant
  hydrogen gas is observed out to $d\sim 500$ kpc, well beyond the
  dark matter halo radius $R_h$ of individual galaxies, with a mean
  covering fraction of $\approx 60$\%.  In contrast, no heavy elements
  are detected at $d\apg 0.7\,R_h$ from either low-mass dwarfs or
  high-mass galaxies.  The lack of detected heavy elements in low-
  and high-ionization states suggests that either there exists a
  chemical enrichment edge at $d\approx 0.7\,R_h$ or gaseous clumps
  giving rise to the observed absorption lines cannot survive at these
  large distances.  Considering all galaxies at $d>R_h$ leads to a
  strict {\it upper limit} for the covering fraction of heavy elements
  of $\approx 3$\% (at a 95\% confidence level) over $d=(1-9)\,R_h$.
  At $d<R_h$, differential covering fraction between low- and
  high-ionization gas is observed, suggesting that the CGM becomes
  progressively more ionized from $d<0.3\,R_h$ to larger distances.
  Comparing CGM absorption observations at low and high redshifts
  shows that at a fixed-fraction of $R_h$ the CGM exhibits stronger
  mean absorption at $z=2.2$ than at $z\sim 0$ and that the
  distinction is most pronounced in low-ionization species traced by
  C\,II and Si\,II absorption lines.  We discuss possible
  pseudo-evolution of the CGM as a result of misrepresentation of halo
  radius, and present a brief discussion on the implications of these
  findings.

\end{abstract}

\begin{keywords}
galaxies:halos -- galaxies: dwarf -- quasars: absorption lines -- intergalactic medium -- survey
\end{keywords}

\section{INTRODUCTION}
 

Circumgalactic space is thought to contain a vast amount of baryons
(e.g.\ Spitzer 1956; Fukugita 2004) that are both multiphase and
dynamic (e.g.\ Mo \& Miralda-Escud\'e 1996, Maller \& Bullock 2004).  In
principle, this circumgalactic gas provides the fuel necessary for
sustaining star formation in galaxies, while at the same time it is
also being replenished with both newly accreted (and presumably
chemically pristine) intergalactic gas and chemically enriched
materials either through gas stripping due to satellite interaction or
through starburst driven winds.  Therefore, circumgalactic space
provides a critical laboratory for studying the baryon cycle that
regulates star formation and galaxy growth.


Over the past two decades, absorption spectroscopy of background QSOs
have provided an effective tool for probing the otherwise unseen,
diffuse gas in halos around distant galaxies.  While each QSO provides
a one-dimensional mapping through each single halo, observing an
ensemble of close galaxy and QSO pairs leads to statistical
measurements of mean halo profiles (i.e.\ absorption strength as a
function of projected distance) averaged over the entire galaxy
sample.  Commonly seen transitions in circumgalactic space include
hydrogen \lya\,$\lambda\,1215$ (e.g.\ Lanzetta \etal\ 1995; Chen
\etal\ 1998, 2001a; Tripp \etal\ 1998; Wakker \& Savage 2009; Steidel
\etal\ 2010; Prochaska \etal\ 2011; Thom \etal\ 2012; Stocke \etal\
2013; Rudie \etal\ 2013), the C\,IV\,$\lambda\lambda\,1548, 1550$
doublet (e.g.\ Chen \etal\ 2001b; Adelberger \etal\ 2005; Steidel
\etal\ 2010), the Mg\,II\,$\lambda$$\lambda\,2796,2803$ doublet (e.g.\
Bowen \etal\ 1995; Chen \& Tinker 2008; Kacprzak \etal\ 2008; Barton
\& Cooke 2009; Chen \etal\ 2010a,b; Gauthier \etal\ 2010; Bordoloi
\etal\ 2011; Werk \etal\ 2013), and the
O\,VI\,$\lambda\lambda\,1031,1036$ doublet (Chen \& Mulchaey 2009;
Wakker \& Savage 2009; Prochaska \etal\ 2011; Tumlinson \etal\ 2011;
Johnson \etal\ 2013).  A general finding from different studies is
that most galaxies are surrounded by chemically enriched gas out to
$\approx 100-300$ kpc in projected distance.  Competing scenarios for
explaining the presence of heavy elements at $> 50$ kpc from
star-forming regions include super galactic winds (e.g.\ Murray \etal\
2011; Booth \etal\ 2013) and tidally disrupted satellites (e.g.\ Wang
1993; Gauthier 2013).

A particularly interesting absorption feature to adopt for probing the
baryon content of galactic halos is the \civ\ doublet.  These doublet
transitions are strong and their rest-frame wavelengths, 1548.20 and
1550.77 \AA, enable uniform surveys of chemically enriched gas in the
optical window over a broad redshift range from redshift $z\approx 1$
to $z\apg 5$.  \civ\ absorbers are found to originate primarily in
photoionized gas of temperature $T\approx 4 \times 10^4$ K with some
contribution from shock heated gas in galaxies and galactic halos
(e.g.\ Rauch \etal\ 1997; Boksenberg \etal\ 2003; Simcoe \etal\ 2004).
Therefore, \civ\ absorption transitions provide an effective tracer of
chemically enriched warm gas in and around star-forming regions.

A number of surveys have been carried out to characterize the
statistical properties of \civ\ absorbers found along random QSO
sightlines (e.g.\ Songaila 2001; Boksenberg \etal\ 2003; Simcoe 2011;
Cooksey \etal\ 2013).  While the shape of the \civ\ equivalent width
frequency distribution function appears to remain the same, the total
cosmic mass density in C$^{3+}$ ions is found to increase from $z=4$
to $z\approx 1.5$ by a factor of 2 (e.g.\ Cooksey \etal\ 2013).  The
increasing \civ\ mass density together with an increasing background
radiation intensity with decreasing redshift (e.g. Haardt \& Madau
2012) indicate an increasing chemical enrichment level with time in
the gas traced by the \civ\ absorption transitions (e.g.\ Oppenheimer
\& Dav\'e 2008).


At the same time, only three studies have been carried out to
characterize extended \civ\ gaseous halos around galaxies, two at
$z\apl 0.5$\footnote{Though we have recently learned that Bordoloi
  \etal\ (2014) has also been conducting a survey of \civ\ absorbers
  in halos around low-redshift galaxies.} and one at $z\approx 2.2$.
A clear understanding has yet to be established.  At $z\approx 0.4$,
Chen \etal\ (2001b) studied the incidence of \civ\ absorbing gas at
projected distances $d\apl 300$ kpc from a sample of 50 galaxies.
These authors reported the presence of a distinct boundary at a
$B$-band luminosity-normalized projected distance of $\hat{d}\equiv
d\times(L_B/L_B^*)^{-0.5}\approx 160$ kpc, beyond which no \civ\
absorbers are detected.  The lack of \civ\ absorption is based on a
subsample of 35 galaxies that occur at $\hat{d}> 160$ kpc from their
background QSO sightlines.  In contrast, 14 of the 15 galaxies at
$\hat{d}<160$ kpc have an associated \civ\ absorber, although the
rest-frame absorption equivalent width, \ewr, exhibits a large scatter
between individual galaxies.  At $z\approx 2.2$, Steidel \etal\ (2010)
examined the mean spatial absorption profiles of 512 galaxies based on
stacked spectra of background galaxies that occur at $d\apl 125$ kpc.
These authors found that the mean \civ\ absorption strength declines
rapidly at projected distances of $\approx 50-100$ kpc.  Comparing the
observations of Chen \etal\ (2001b) and Steidel \etal\ (2010) yielded
little distinction in the \civ-traced circumgalactic medium (CGM) at
low and high redshifts (Chen 2012).  Both the luminosity-normalized
spatial extent and mean absorption equivalent width of the CGM around
galaxies of comparable mass (but with very different on-going star
formation rate) have changed little over the redshift interval
$z=0.4-2.2$, although there exists a large scatter in \ewr\ in the
low-redshift study.  Nevertheless, a lack of variation in the spatial
profile of the chemically enriched CGM between two distinct epochs
poses a serious challenge to the theoretical models of gas flows
around galaxies (e.g.\ Hummels \etal\ 2013; Ford \etal\ 2013a,b).

On the other hand, Borthakur \etal\ (2013) targeted a sample of 20
galaxies at $z<0.2$ and were able to constrain the incidence of \civ\
absorbing gas for 17 of these galaxies.  Of the 17 galaxies studied,
eight occur at $\hat{d}<160$ kpc and nine at larger distances.  These
authors detected \civ\ absorbers for three of the eight galaxies at
$\hat{d}<160$ kpc and three of the nine galaxies at $\hat{d}>160$ kpc
in their sample.  The finding of a flat rate of \civ\ incidence with
increasing projected distance is intriguing and appears to be in stark
contrast to the earlier finding of Chen \etal\ (2001b).  While
Borthakur \etal\ attributed the detected \civ\ absorbers to starburst
activity in the associated galaxies, it is not straightforward to
reconcile the discrepant trend found for extended \civ\ halos between
the Chen \etal\ and Borthakur \etal\ samples.

To address the discrepant trend found for extended \civ\ halos around
galaxies at low redshifts and to improve the empirical understanding
of how the chemically enriched CGM evolves with time, we have searched
public archives to find spectroscopically identified galaxies at small
projected distance to a UV bright, background QSO for which
high-quality echelle/echellette spectra are already available in the
{\it Hubble Space Telescope} (HST) data archive.  Our search has
yielded the first large sample of $\sim 300$ close QSO and galaxy
pairs that enables a systematic study of extended \civ\ halos over the
projected distance interval of $d = 0-500$ kpc.  As described below,
this unique sample allows us to place unprecedented limits on the
extent of chemical enrichment around low-redshift galaxies based on
observations of not only the \civ\ absorption doublets but also a
whole host of ionic transitions from low- and intermediate-ionization
states.

Recently, Tumlinson \etal\ (2013) carried out a large program to use
the Cosmic Origins Spectrograph (COS; Green et al.\ 2012) on board HST
to study the CGM at $d<150$ kpc of 44 $L_*$ galaxies at $z=0.15-0.35$.
This program, known as the ``COS-Halos'' survey, was designed to map
the multi-phase CGM using O\,VI and other metal-line diagnostics
(Tumlinson \etal\ 2011; Werk \etal\ 2013).  The low-redshift cut at
$z=0.15$ was dictated by the requirement of detecting the O\,VI
doublet using COS and the FUV channel.  However, the spectral coverage
of the COS FUV gratings misses the C\,IV doublet at $z\apg 0.15$, and
therefore \civ\ transitions are not included in the COS-Halos analysis
(e.g.\ Werk \etal\ 2013).

Our study differs from the COS-Halos program in two fundamental
aspects.  First, our search is designed to probe the \civ\ absorption
in galactic halos for comparisons with high-redshift studies.
Second, because fainter galaxies are also more numerous, it is more
likely to find a faint foreground galaxy near a QSO sightline than a
luminous galaxy (see the discussion is \S\ 2).  As a result,
low-luminosity ($<0.1\,L_*$) and low-mass ($<0.1\,M_{\rm star}^*$)
dwarf galaxies contribute to a large fraction ($\approx 50$\%) of our
random QSO and galaxy pair sample.  Therefore, {\it our study
  complements the COS-Halos effort both in terms of the mass regime of
  the galaxies and in terms of the ionization state of the CGM}.


Here we report initial findings for the low-redshift CGM based on a
sample of 195 galaxies for which constraints on the stellar mass are
available.  The available stellar mass estimate of each galaxy allows
us to infer the total mass of the dark matter in which the galaxy
resides (e.g.\ Behroozi et al.\ 2013; Kravtsov et al.\ 2014). This in
turn allows us to assess the intrinsic size differences between
different galaxies based on the estimated dark matter halo radius from
a standard halo model.  While the CGM is expected to be regulated by
various complicated physical processes, including gravitational motion
and star formation feedback, addressing the effects of all these
processes at once is challenging.  As a first step toward establishing
a clear understanding of the processes that drive the observed CGM
properties, we have designed our study here to account for the
intrinsic size differences (driven by the gravitational potential) of
individual galactic halos by normalizing the observed projected
distances with the underlying halo radius.  We expect that the results
will facilitate future studies that consider the effect of additional
processes such as star formation feedback.

This paper is organized as follows.  In Section 2, we describe the
selection criteria for establishing the close galaxy--QSO pair sample
and summarize the properties of the galaxy members in the pair sample.
In Section 3, we describe the data reduction and analysis procedures
of the absorption spectra of the QSOs.  In Section 4, we examine the
correlation between the strengths of different ionic transitions and
galaxy properties, and compare the results of our study with previous
findings.  Finally, we discuss the implication of our results in
Section 5.  Throughout the paper, we adopt the standard $\Lambda$
cosmology, $\Omega_\Lambda = 0.7$, $\Omega_M = 0.3$ with a Hubble
constant $H_0 = 70\,\kms\,{\rm Mpc}^{-1}$.


\section{A Public Sample of Galaxy--QSO Pairs}

We have assembled a large sample ($n_{\rm gal} \sim\ 300$) of
spectroscopically-identified galaxies that occur at small projected
distances $\apl 500$ kpc from the sightline of a background UV bright
QSO.  This galaxy sample is ideal for characterizing extended gaseous
halos around galaxies based on the absorption features imprinted in
the spectra of the background QSOs.

We first searched the HST archive for QSOs that have been observed
with either the Space Telescope Imaging Spectrograph (STIS; Woodgate
et al.\ 1998) or the Cosmic Origins Spectrograph (COS; Green et al.\
2012) as of Cycle 20.  This search yielded a sample of 150 QSOs at $
z_{\rm{QSO}} = 0.058-1.476$.  Next, we searched for spectroscopically
identified galaxies in public archives, including the Nearby General
Catalog (NGC; Sinnott 1988), the Sloan Digital Sky Survey (SDSS; York
\etal\ 2000) spectroscopic galaxy sample, the Two Micron All Sky
Survey (2MASS) galaxies (Huchra et al.\ 2012), and the Two Degree
Field Galaxy Redshift Survey (2dFGRS; Colless et al.\ 2001).  A galaxy
and a background QSO were considered a suitable pair if they satisfied
the following criteria.  (1) At the redshift of the galaxy, the
physical projected distance $d$ between the galaxy and the QSO
sightline is $d \leq 500$ kpc.  (2) The galaxy must be at
$|\Delta\,v|> 10000$ \kms\ below the redshift of the QSO to avoid
ambiguity between foreground absorbers and those originating in QSO
outflows (e.g.\ Wild \etal\ 2008).  The maximum separation limit of
500 kpc is sufficiently large to encompass the gaseous halos of $L_*$
galaxies known empirically (e.g.\ Chen et al.\ 1998, 2001a; Gauthier
et al.\ 2011; Prochaska et al.\ 2011; Rudie \etal\ 2012).  It is also
motivated by theoretical expectations of extended halo gas out to
$\sim 300$ kpc in radius from star-forming regions (e.g.\ Ford et al.\
2013a).  Therefore, including pairs separated by greater than 300 kpc
allows us to explore the presence of chemically enriched gas beyond
the fiducial virial radius of dark matter halos.

Finally, we formed an ``isolated'' galaxy and QSO pair sample by
considering only galaxies without close neighbors, in order to
minimize the ambiguity in associating an absorber with a galaxy.
Previous studies have also shown that galaxies in group environments
tend to have more extended halo gas than isolated galaxies that
results in a larger scatter in the observed spatial distribution of
absorption strength with projected distance (Chen et al.\ 2010a,
Bordoloi et al.\ 2011a, and Gauthier \& Chen 2011).  In our study, a
galaxy was considered an ``isolated'' one, if no other galaxies have
been spectroscopically identified within projected separation of
$d=500$ kpc and velocity separation of $|\Delta\,v|=500$ \kms\ were
present.  These criteria are driven by the model expectation that an
$L_*$ galaxy at $z=0$ has a virial radius of $\approx 300$ kpc and
maximum circular velocity of $\approx 200$ \kms.
We chosen a more conservative limit to allow for uncertainties in
galaxy redshifts.  

\begin{figure}
	\centering
	\includegraphics[scale =0.45]{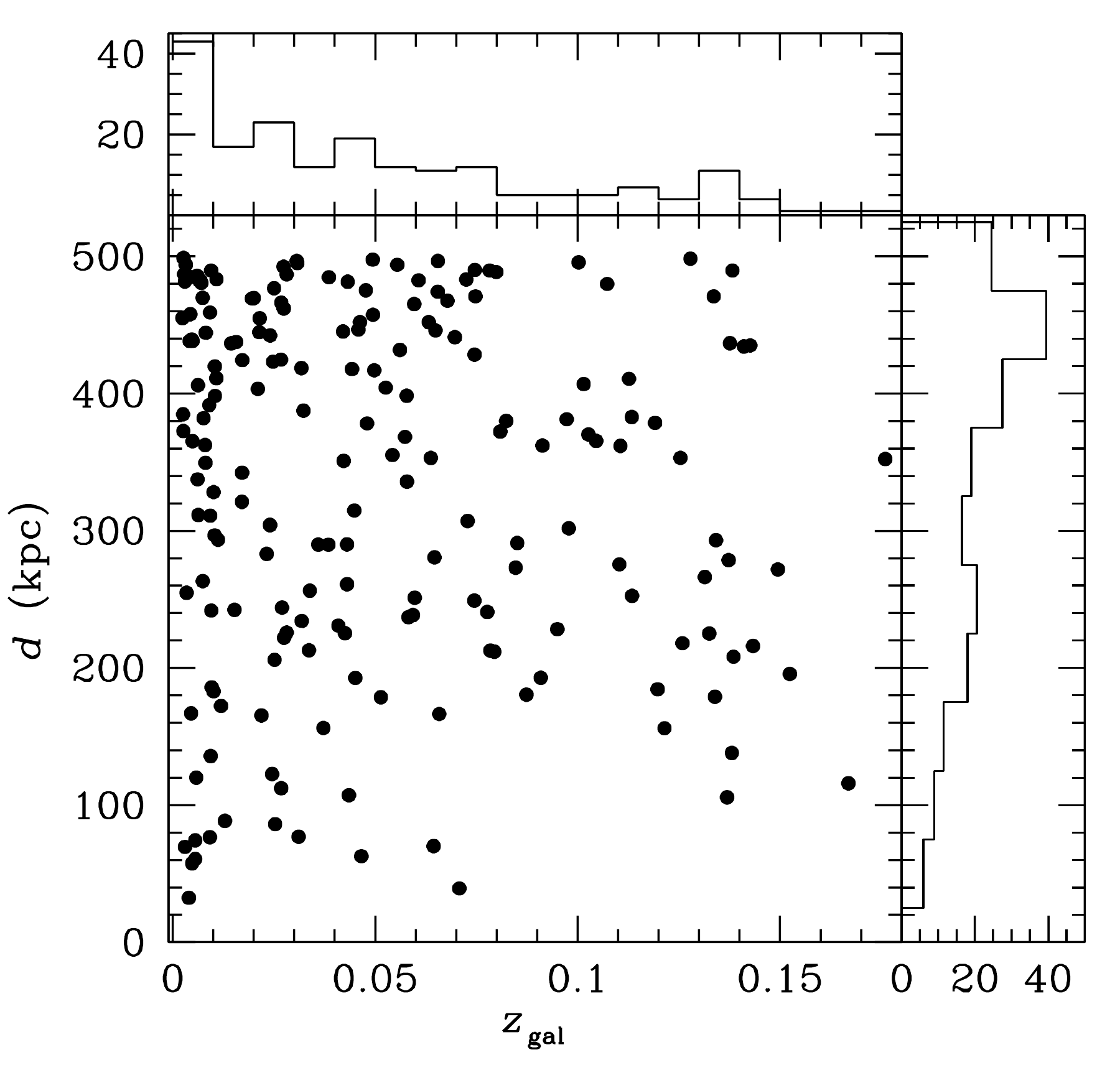}
	\caption{Distribution of projected distance versus redshift
          for the 195 galaxies (\S\ 2).  The top (right) panel
          displays the redshift (impact parameter) histograms. }
\end{figure}

 
Our search criteria of suitable galaxy and QSO pairs yielded a final
sample of 213 ``isolated'' galaxies that occur at $d \apl 500$ kpc
from the sightline of a background QSO for studies of the low-redshift
CGM.  Excluding 18 galaxies that are in the COS-Dwarfs program
(Bordoloi \etal\ 2014, in preparation) led to a total of 195 galaxies
in our sample.  The redshifts of the galaxies range from $z = 0.002$
to $z = 0.176$ with a median of $\langle\,z\,\rangle = 0.041 $, and
the projected distances of the QSOs range from $d\approx 32$ kpc to
$d\approx 500$ kpc with a median of $\langle\,d\,\rangle=362$ kpc.
The projected distance versus redshift distribution of the galaxies
sample is shown in Figure 1, in which the top and right panels display
the histograms in redshift and in projected distance, respectively.

To facilitate a detailed investigation of how the observed CGM
absorption properties depend on galaxy properties, we also made use of
available measurements of rest-frame UV luminosities and stellar mass
from the NASA--Sloan Atlas\footnote{http://nsatlas.org/} (Blanton
\etal\ 2005; Blanton \etal\ 2011).  This public atlas contains 145,155
galaxies with known redshifts at $z\apl 0.05$.  For each galaxy, it
includes optical and UV photometric measurements from the SDSS Data
Release 8 (Blanton \etal\ 2011) and GALEX Data Release 6 (Schiminovich
\etal\ 2007), as well as derived quantities such as stellar mass
($M_{\rm star}$) and rest-frame UV and optical absolute magnitudes
computed by the K-correct code (Blanton \& Roweis 2007).  The
rest-frame UV absolute magnitudes allows us to estimate an
unobscured star formation rate (SFR) for each galaxy based on the
calibration coefficient of Kennicutt \& Evans (2012).  The available
stellar mass estimate of each galaxy allows us to infer the total mass
of the dark matter halo ($M_h$) in which the galaxy resides (e.g.\
Behroozi \etal\ 2013; Kravtsov \etal\ 2014).  This in turn allows us
to assess the intrinsic size differences between different galaxies
based on the estimated dark matter halo radius $R_h$ from a standard
halo model (see \S\ 4 for a more detailed discussion).

\begin{figure}
	\centering
	\includegraphics[scale=0.45]{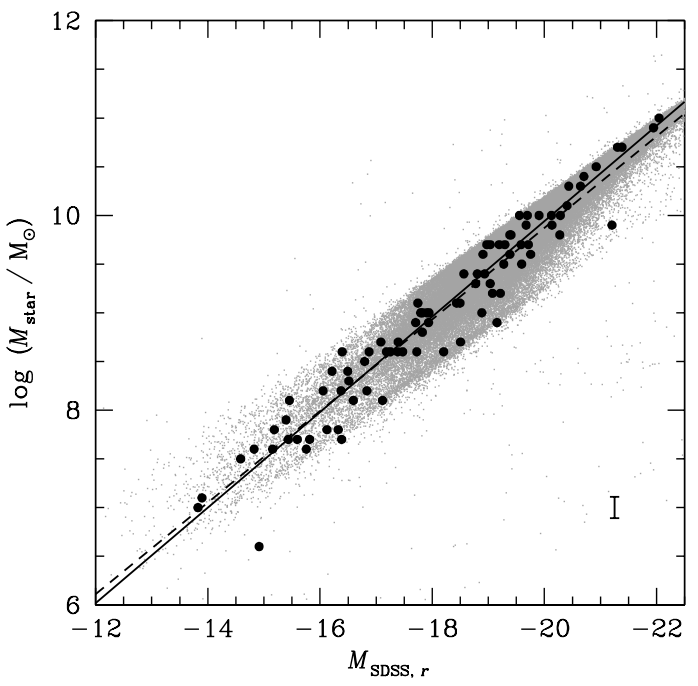}
	\caption{Observed correlation between stellar mass $M_{\rm
            star}$ and rest-frame $r$-band absolute magnitude,
          $M_{{\rm SDSS},r}$ for 111 overlapping galaxies that are
          both in our pair sample and in the NASA--Sloan Atlas (dark
          solid points).  The solid line indicates the best-fit model
          with $\log\,M_{\rm star}=0.14-0.49\,M_r$ and a 1-$\sigma$
          scatter of $\Delta\,\log\,M_{\rm star}=0.21$ (error bar in
          the lower-right corner).  The full NASA--Sloan catalog,
          which has $>100$k galaxies (grey points), is also included
          to contrast these 111 overlapping galaxies.  The full sample
          displays a consistent mean relation (dashed line) and
          scatter between $M_{\rm star}$ and $M_r$.  The consistent
          distribution of the two samples strongly supports that the
          mean relation derived using the subsample of 111 galaxies is
          representative of the general galaxy population at low
          redshifts.}
\end{figure}

Cross-matching the ``isolated'' galaxy sample with the NASA--Sloan
Atlas resulted in 116 overlapping galaxies.  We visually inspected the
optical images of each galaxy and removed five objects due to
erroneous photometry.  This exercise resulted in a sample of 111
galaxies with accurate $M_{\rm star}$ and star formation rate (SFR)
measurements available from the NASA--Sloan Atlas.  However, to
utilize the full ``isolated'' galaxy sample for investigating how the
observed CGM absorption properties depend on galaxy properties, we
needed to estimate the stellar mass for each of the remaining 100
galaxies based on available photometric data.  We used the subsample
of 111 galaxies as a guide.  Comparing the stellar mass provided by
the NASA--Sloan Atlas with intrinsic luminosities in different
bandpasses revealed a tight correlation between $M_{\rm star}$ and
rest-frame absolution $r$-band magnitude, $M_{{\rm SDSS},r}$, over
four orders of magnitude in $M_{\rm star}$ (Figure 2).  Based a linear
regression analysis, we find that the $M_{\rm star}$ vs.\ $M_{{\rm SDSS},r}$
correlation is best characterized by
\begin{equation}
\log\,M_{\rm star}=0.14-0.49\,M_{{\rm SDSS},r}
\end{equation}
with a 1-$\sigma$ scatter of $\Delta\,\log\,M_{\rm star}=0.21$.
Equation (1) enables us to obtain a stellar mass estimate and
associated uncertainty for galaxies based on $M_{{\rm SDSS},r}$.  In
Figure 2, we also include the entire NASA--Sloan catalog, which has
$>100$k galaxies (grey points), to contrast the 111 galaxies adopted
for deriving Equation (1).  The comparison shows that the 111
overlapping galaxies and the full NASA--Sloan catalog share a
consistent mean relation and scatter between $M_{\rm star}$ and
$M_{{\rm SDSS},r}$, particularly at the high-mass end with $M_{\rm
  star}>10^{10}\,M_\odot$.  Equation (1) is therefore representative
of the general galaxy population at low redshifts.

\begin{figure}
	\centering
	\includegraphics[scale =0.42]{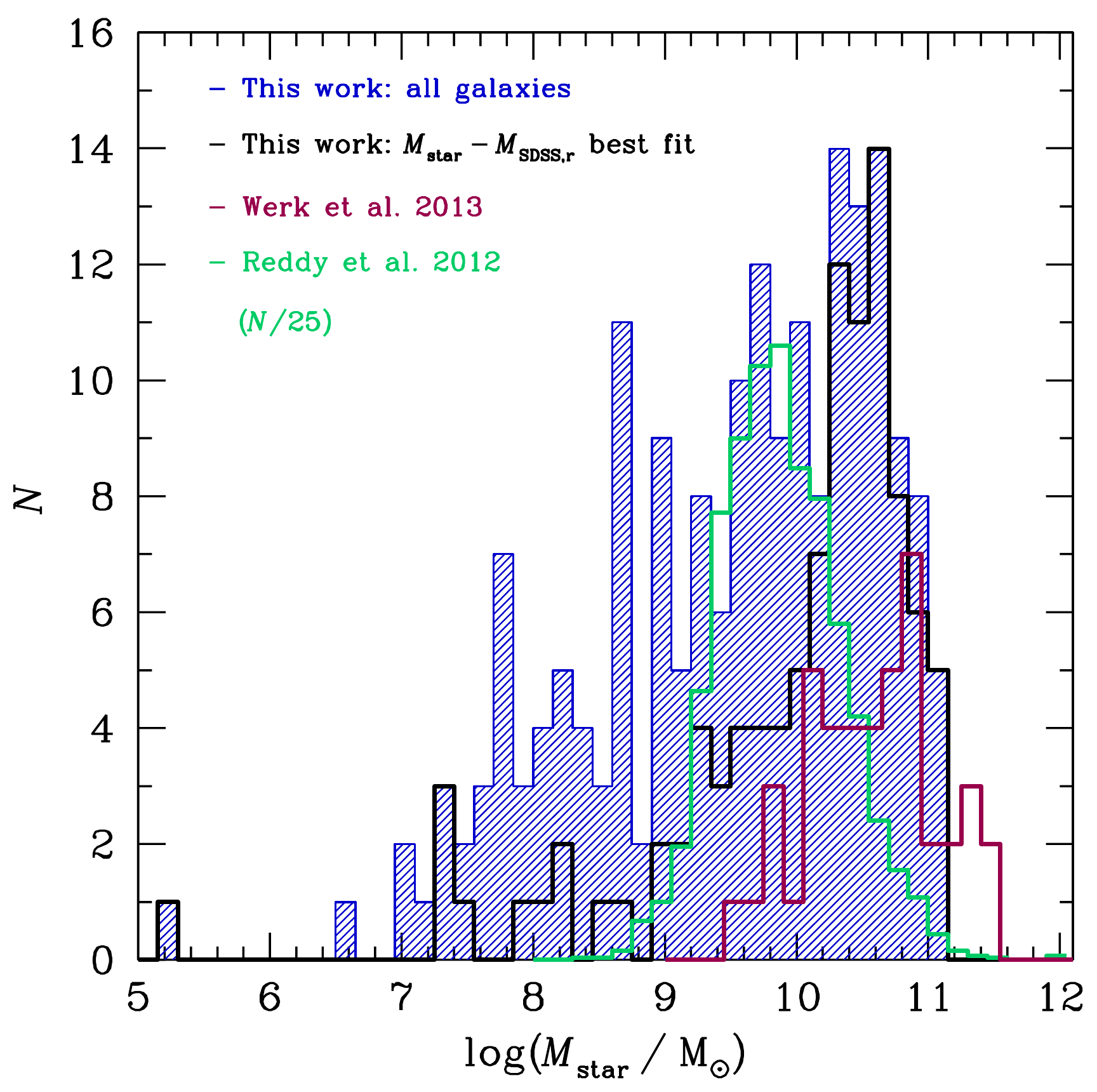}
	\caption{Stellar mass distribution of our galaxy sample (blue
          shaded histogram).  The median mass is $\langle\,M_{\rm
            star}\,\rangle=5\times 10^{9}\,{\rm M}_\odot$, in
          comparison to the characteristic mass $M_{\rm
            star}^*=5\times 10^{10}\,{\rm M}_\odot$ for the field
          galaxies (Baldry \etal\ 2012; Muzzin \etal\ 2013). Our
          galaxy sample has a mass distribution that encompasses
          previous CGM studies at both low and high
          redshifts. Specifically, the COS-Halos sample at $z=0.1-0.4$
          (Werk \etal\ 2013; red open histogram) has a median stellar
          mass of $4\times 10^{10}\,{\rm M}_\odot$ and the starburst
          sample at $z=2.2$ from Reddy \etal\ (2012; green open
          histogram) has a median stellar mass of $7.9\times
          10^{9}\,{\rm M}_\odot$.  The number counts of the Reddy
          \etal\ sample has been reduced by a factor of 25 for
          presentation purpose.}
          
\end{figure}


We present the distribution of stellar masses of our galaxies in
Figure 3.  The blue shaded histogram shows the stellar mass
distribution for the full sample of 195 galaxies, while the black open
histogram shows the stellar mass distribution for the subsample of 102
galaxies with $M_{\rm star}$ inferred from $M_r$ following Equation
(1).  Figure 3 shows that the galaxies in the full sample span a wide
range in stellar mass, from $M_{\rm star} = 1.5\times 10^5\,{\rm
  M}_\odot$ to $M_{\rm star} = 1.4 \times 10^{11}\,{\rm M}_\odot$,
with a median $\langle\,M_{\rm star}\,\rangle=5.0\times 10^{9}\,{\rm
  M}_\odot$, which is $\approx 0.1\,M_{\rm star}^*$ (Baldry \etal\
2012; Muzzin \etal\ 2013).  Comparing the blue shaded and black open
histograms shows that including galaxies with $M_{\rm star}$ inferred
from Equation (1) allows us to expand our CGM study to include
higher-mass galaxies, covering a mass range that overlaps with the
COS-Halos sample at $z=0.1-0.4$ (Werk \etal\ 2013; red open histogram)
and the starburst sample at $z=2.2$ from Reddy \etal\ (2012; green
open histogram)\footnote{We note that while different galaxy samples
  overlap in the stellar mass range, there are strong distinctions in
  SFR and in the knowledge of environments and redshift precisions.
  Specifically, the starburst sample at $z=2.2$ have a mean SFR of
  $\sim 30\,M_\odot\,{\rm yr}^{-1}$ (e.g.\ Erb \etal\ 2006) and
  low-redshift galaxies have on average unobscured SFR of $\sim
  0.1\,M_\odot\,{\rm yr}^{-1}$.  In addition, while our sample only
  includes galaxies in an isolated environment, the environments of
  the high-redshift starburst sample are unknown.}.

In addition, examining the SFR versus $M_{\rm star}$ correlation helps
to characterize the galaxies in our pair sample in the context of the
general galaxy population.  Considering the 111 galaxies with both
$M_{\rm star}$ and rest-frame UV photometric measurements from the
NASA--Sloan Atlas, we present in Figure 4 the specific Star Formation
Rate (sSFR; SFR per unit stellar mass) and stellar mass of these
galaxies (solid points) in comparison to the mean relations of
star-forming main sequence galaxies (blue line) and passive, red
galaxies (red) from Schiminovich et al.\ (2010).  Recall that we
measured an unobscured SFR for each galaxy based on the observed NUV
flux without correcting for possible dust extinction.  Therefore, our
measurements likely represent lower limits to the intrinsic SFR.  The
tilt in the distribution of our galaxies relative to the mean relation
of star-forming main-sequence galaxies can be explained by the
observed correlation between dust extinction and stellar mass, namely
more massive galaxies on average exhibit higher degrees of dust
extinction (e.g.\ Zahid \etal\ 2013).  It is clear that the majority
of the galaxies in our sample are consistent with being star-forming
main-sequence galaxies with a small fraction found in the red-sequence
regime.

We note that while the stellar mass derived using the K-correct code
may be systematically overestimated by 0.1 dex in the low mass regime
of $M_{\rm star}\apl 10^{10}\,{\rm M}_\odot$ (e.g.\ Moustakas \etal\
2013), such error translates to a systematic uncertainty of $\approx
0.06$ dex in $M_h$, well within the scatter in the
stellar-mass-to-halo-mass relation of Behroozi \etal\ (2013).  The
resulting uncertainty in $R_h$ is then negligible.

Of the 195 galaxies in our pair sample, 12 occur within 100 kpc in
projected distance from a background QSO sightline.  We summarize the
properties of the galaxy sample in Table 1, which lists for each
galaxy the identification, right ascension (RA) and declination (Dec),
spectroscopic redshift ($z_{\rm gal}$), unobscured SFR (when
available), stellar mass ($M_{\rm star}$), halo mass ($M_h$), halo
radius ($R_h$), and absolute $r$-band magnitude $M_{{\rm SDSS},r}$. 

\begin{figure}
	\centering
	\includegraphics[scale=0.33]{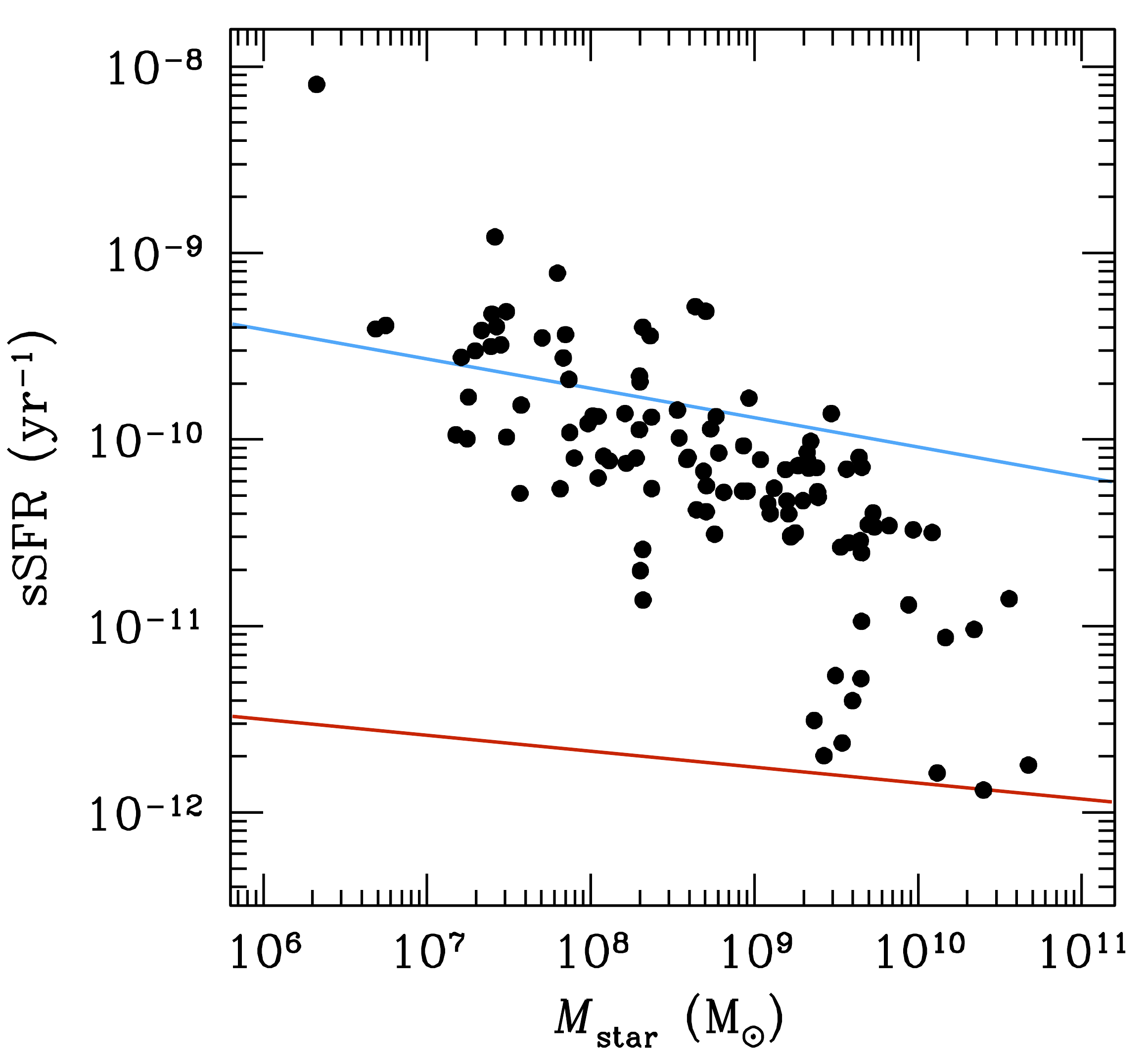}
	\caption{Distribution of specific star formation rate (sSFR)
          and stellar mass based on a subsample of our galaxies with
          available UV photometric from the NASA-Sloan Atlas (solid
          points).  SFR is derived based on the observed NUV flux from
          GALEX and therefore represents a lower limit to the
          intrinsic value.  The blue (red) line marks the star-forming
          main-sequence (passive, red) galaxies from Schiminovich et
          al.\ (2010).  The tilt in the distribution of our galaxies
          relative to the mean relation of star-forming main-sequence
          galaxies can be explained by more massive galaxies showing
          higher degrees of dust extinction (e.g.\ Zahid \etal\ 2013).
          Nevertheless, the majority of our galaxies are consistent
          with being star-forming main-sequence galaxies, and a small
          fraction are found in the red-sequence regime.}
\end{figure}


\section{QSO UV Spectroscopy}

\begin{figure*}
\begin{center}
\includegraphics[scale=0.6]{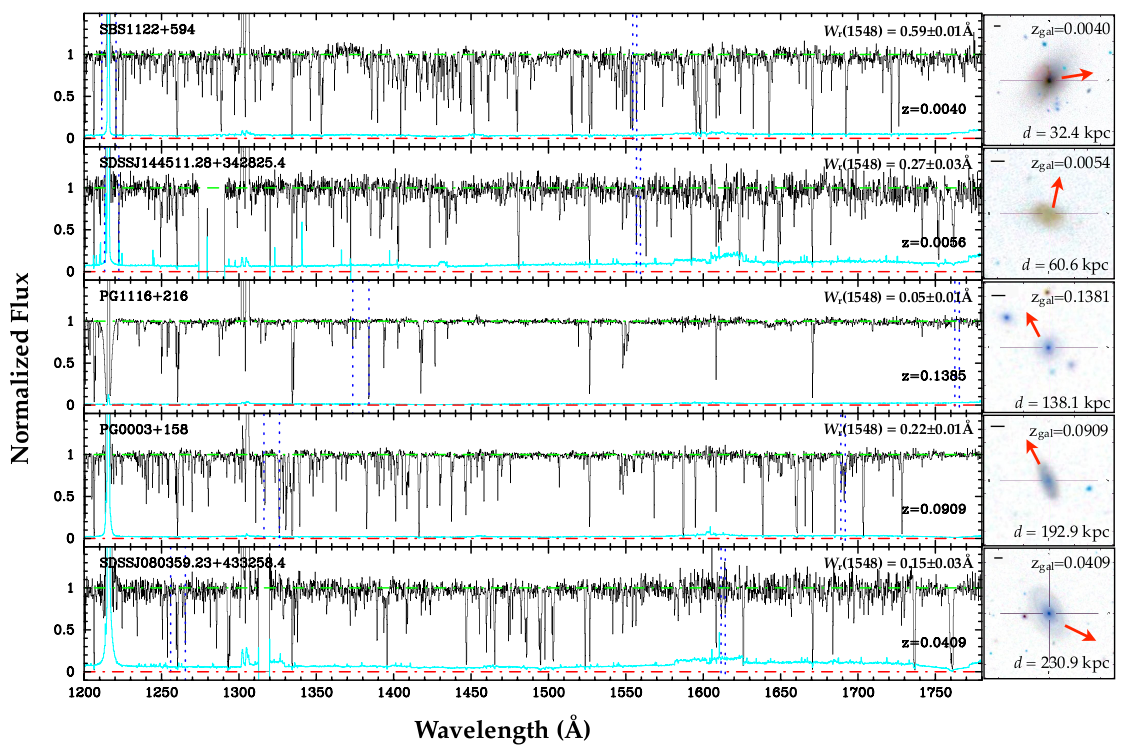}
\caption{Examples of galaxies with associated \civ\ absorbers from our
  search.  For each field, we present the QSO spectrum in the left
  panel and the corresponding SDSS image centered at the galaxy in the
  right panel.  In each panel, the direction to the QSO sightline is
  indicated by the arrow, and the horizontal bar in the upper-left
  corner indicates an angular scale of $5''$.  The projected
  separation between the QSO sightline and the galaxy is shown in the
  lower-right corner.  Absorption features due to
  Si\,III\,$\lambda\,1206$, \lya\,$\lambda\,1215$, and
  \civ\,$\lambda\lambda\,1548,1550$ at the redshifts of the galaxies
  are indicated by blue, dotted lines in the left panels.  The
  absorber redshift is shown in the lower right corner of each
  spectrum.}
\end{center}
\end{figure*}

Our galaxy sample contains 195 ``isolated'' galaxies that occur near
the sightlines of 96 independent background QSOs with high-quality,
far-ultraviolet echellette spectra available in the HST data archive.
Of the 96 QSOs, 13 were observed using STIS and the E140M grating and
83 were observed using COS and the G130M/G160M gratings.  STIS and the
E140M grating offers a contiguous spectral coverage from $\approx
1140$ \AA\ to $\approx 1740$ \AA\ with a full-width-at-half-maximum
resolution of ${\rm FWHM}\approx 7$ \kms, while COS with G130M and
G160M at different central wavelengths offers a nearly contiguous
spectral coverage from $\approx 1150$ \AA\ to $\lambda\approx 1770$
\AA\ with a spectral resolution of ${\rm FWHM}\approx 16$ \kms.

Individual flux-calibrated spectra from STIS were retrieved from the
MAST server.  For each QSO, individual echelle orders were
continuum-normalized and co-added to form a single, stacked spectrum
using our own software.  The continuum was determined using a
low-order polynomial fit to spectral regions that are free of strong
absorption features.  The mean signal-to-noise ($S/N$) of the combined
echelle spectra of 13 QSOs have on average $S/N>7$ over the majority
of the spectral regions covered by the data.

Individual one-dimensional COS spectra were retrieved from the HST
archive and combined using our own software.  In summary, individual
spectra were first aligned using common absorption lines along each
sightline, including both intervening absorbers at $z>0$ and known
Milky Way features, such as Si\,III\,$\lambda\,1206$ and
C\,II\,$\lambda\,1334$ in the G130M data and Si\,II\,$\lambda\,1526$
and Al\,II\,$\lambda\,1670$ in the G160M data.  In a few sightlines,
the pipeline generated individual spectra exhibit wavelength
calibration errors that vary with wavelength, from
$\Delta\lambda\approx -0.06$ \AA\ at $\lambda=1160$ \AA\ to
$\Delta\lambda\approx +0.06$ \AA\ at $\lambda\approx 1600$ \AA.  We
correct for the wavelength calibration errors by employing a low-order
polynomial fit to the observed wavelength-dependent shift in
individual spectral segment.  Then the individual wavelength-corrected
spectra were re-binned to a pixel resolution of $\delta\,v=7.5$ \kms\
and coadded into a single combined spectrum using our own co-addition
code.
The final processed and combined spectra have a median signal-to-noise
of $S/N \approx 2 - 26$.  A summary of the STIS and COS observations
for each QSO is presented in Table 2, which lists from columns (1)
through (5) the name, RA, Dec, emission redshift of the QSO, and the
program ID under which the data were taken.  We also list in columns
(6) and (7) of Table 2 the median $S/N$ of the G130M and G160M spectra
from COS and in column (8) the median $S/N$ of the STIS/E140M spectra.

To facilitate absorption-line measurements, we also continuum
normalized each QSO spectrum using a continuum model determined from a
low-order polynomial fit to spectral regions that were free of strong
narrow line features.  The continuum normalization included strong
damping wings in each QSO spectrum due to either the Milky Way \lya\
absorption or intervening damped \lya\ absorbers.
We did not attempt to fit strong emission features such as the
geocoronal \lya\, and O\,I\,$\lambda\lambda\,1302, 1304$ lines.  These
narrow emission lines were instead manually masked and were not
included in spectral regions for identifying absorption features.

\section{Analysis}

The process described in \S\ 3 yielded high-quality,
continuum-normalized UV echellette spectra of 96 QSOs.  These UV
spectra allow us to investigate the baryon content of the
circumgalactic space within 500 kpc in projected distance of 195
independent, foreground galaxies based on the presence/absence of the
corresponding absorption-line features, in particular the \civ\
doublet.  In this section, we describe absorption-line constraints of
individual galactic halos.  In addition, we present stacked QSO
spectra at the rest-frame of the galaxies, and improved
absorption-line constraints afforded by the stacked spectra.  We show
that with much improved $S/N$, these stacked spectra allow us to place
unprecedented limits for the absorption properties of gaseous halos
around galaxies.

\subsection{Absorption Constraints of Circumgalactic Baryons}


\begin{figure*}
\begin{center}
\subfigure{\includegraphics[scale =0.45]{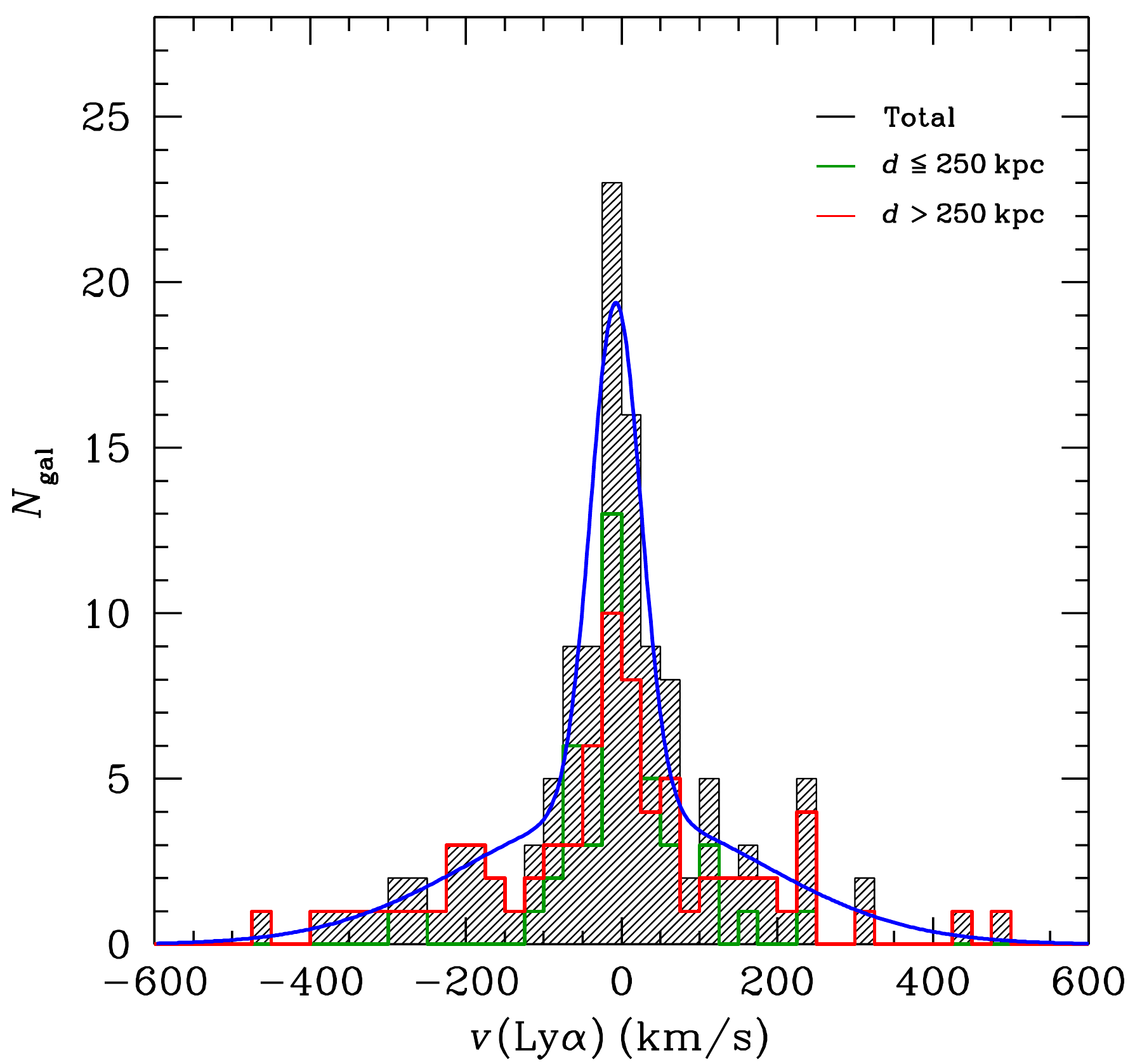}}
\subfigure{\includegraphics[scale =0.45]{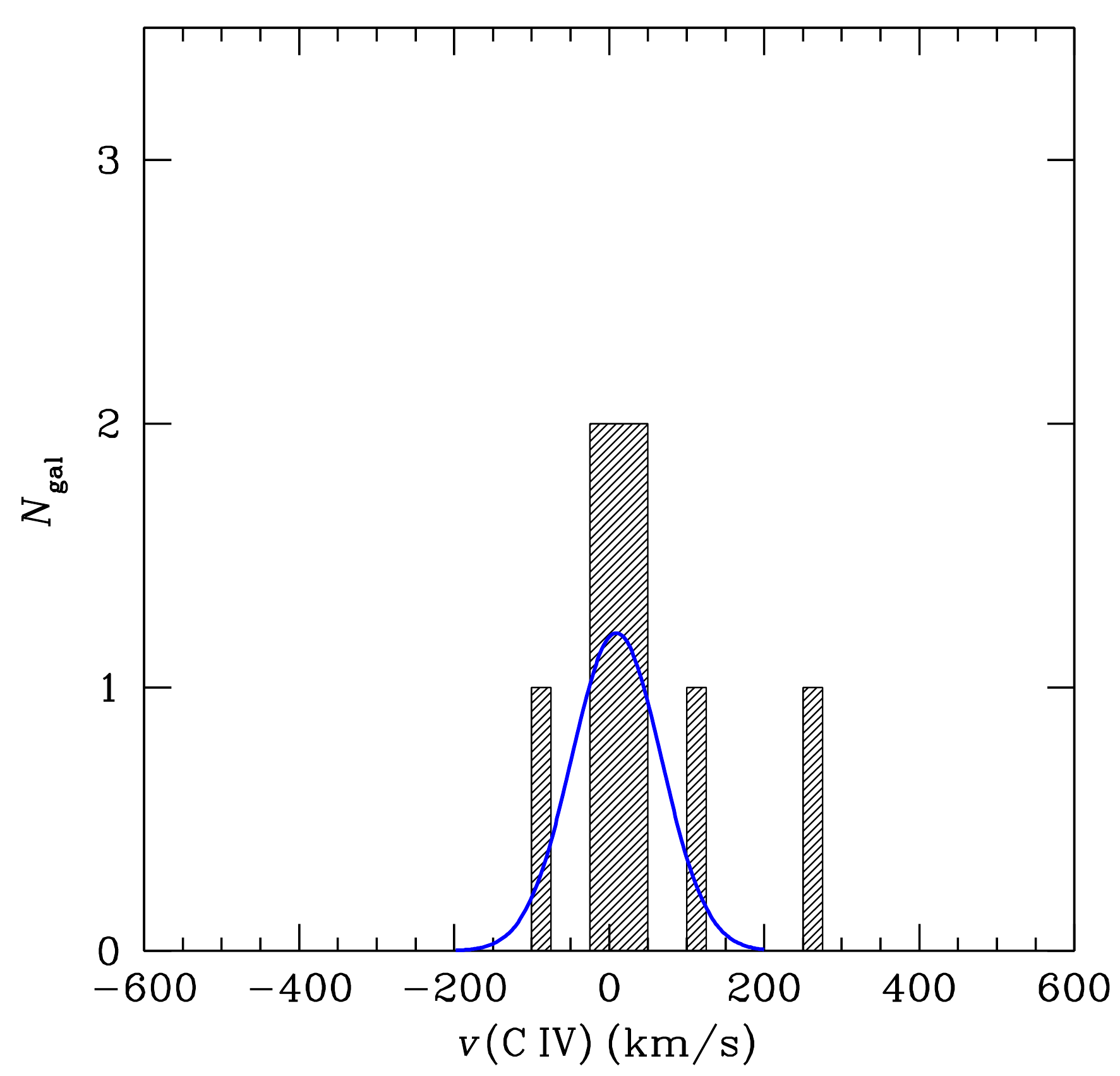}}
\end{center}
\caption{Relative velocity distributions of \lya\ ({\it left}) and
  \civ\ ({\it right}) absorbers with respect to the systemic redshift
  of their associated galaxies in our pair sample.  We detect
  associated \lya\ absorbers in 125 of the 195 galaxies searched, and
  associated \civ\ absorbing gas in nine galaxies.  The left panel
  shows that the velocity distributions of all \lya\ absorbers
  detected around galaxies at $d<500$ kpc (shaded histogram), around
  galaxies at $d<250$ kpc (green open histogram), and galaxies at
  $d>250$ kpc.  Considering the full sample requires a double Gaussian
  profile to characterize the velocity distribution with a narrow
  component centered at $\langle\,v({\rm Ly\alpha})\,\rangle = -7 $
  \kms\ and dispersion of $\sigma_{v}( {\rm Ly\alpha}) = 32$ \kms\ and
  a broad component centered at $\langle\,v({\rm Ly\alpha})\,\rangle =
  -8$ \kms\ and dispersion of $\sigma_{v}({\rm Ly\alpha}) = 188$ \kms\
  (the solid curve).  In contrast, the velocity distribution of \civ\
  absorbing gas around galaxies can be characterized by a single
  Gaussian distribution of $\langle\,v({\rm CIV})\,\rangle = 9$
  \kms\ and $\sigma_{v}({\rm CIV}) = 58$ \kms, excluding the outlier
  at $v({\rm CIV})\approx +260$ \kms.}

\end{figure*}

In order to obtain robust constraints for the baryon content in halos
around our sample galaxies, we first identify and mask contaminating
features associated with other strong \lya\ absorbers at $z\le z_{\rm
  QSO}$ along each QSO sightline, including higher-order Lyman series
and ionic absorption lines.  Then for each of the 195 galaxies, we
search in the associated QSO spectra for the corresponding Ly$\alpha$
absorption line within the velocity range of $|v| \le 500$ \kms\ from
the systemic redshifts of the galaxies.  The allowed large velocity
interval is guided by previous observations (e.g.\ Lanzetta \etal\
1995) which show a velocity dispersion of $\sim 190$ \kms\ between
galaxies and their associated \lya\ absorbers.  For each detected
\lya\ absorber, we further examine whether associated ionic
transitions, such as \civ, C\,II, Si\,IV, S\,III, and Si\,II, are also
present in the spectral range covered by the STIS/COS data.
Absorption transitions considered in our study are summarized in Table
3, together with those considered by Werk \etal\ (2013) and Steidel
\etal\ (2010) for comparison.  We also include in Table 3 the sample
size, redshift range, and stellar mass range of each galaxy sample for
reference.

We detect associated \lya\ absorbers around 125 of the 195 galaxies in
our survey sample, and associated \civ\ absorbing gas in nine
galaxies.  Roughly 30\% of the detected \lya\ absorbers show multiple
components within a velocity interval of 500 \kms\ and only one of the
nine detected \civ\ displays an additional secondary component at
$\Delta\,v=-87$ \kms.  No \lya\ (\civ) measurements can be made for 27
(35) galaxies due to the presence of contaminating features.  This
leaves 43 (151) galaxies that show no trace of \lya\ (\civ) absorbing
gas at $d \le 500$ kpc and $|v|\le 500$ \kms.  Examples of the
galaxies in our sample that display associated \civ\ absorbers are
presented in Figure 5, which also highlights the presence of
associated Si\,III\,$\lambda\,1206$ and \lya\,$\lambda\,1215$ in blue
dotted lines.

To characterize the absorber properties, we measure the rest-frame
absorption equivalent widths for all the observed transitions.  We
focus our current analysis on absorption equivalent width measurements
in order to make direct comparisons with previous studies.  A more
detailed profile analysis is deferred to a future paper (Liang \etal\
2014, in preparation).  In cases where an absorption transition is not
detected, we measure a 2-$\sigma$ upper limit to the rest-frame
absorption equivalent width over a spectral window that corresponds to
the median line width (FWHM) of individual, observed components of the
targeted transition from the full sample.  For galaxies without
associated \lya\ absorption, the upper limits of the underlying \lya\
and \civ\ absorption strengths are estimated at the systemic redshifts
of the galaxies.  For galaxies with associated \lya\ absorption but no
detectable \civ\ features, the upper limits of the underlying \civ\
absorption strengths are estimated at the systemic velocities of the
\lya\ absorbers which are the velocity centroids of the strongest
components determined by the fitting routine described below.

To measure the line widths and velocity centroids of each detected
absorber, we model the observed absorption profile as a collection of
individual Gaussian components and consider only the minimum number of
components necessary to fully describe the observed absorption
features.  The fitting routine returns for each component the velocity
centroid, the 1-$\sigma$ line width, and the maximum absorption depth.
Considering all the individual \lya\ components observed in 125
galaxies, we measure a median component line width of ${\rm
  FWHM}(\lya)=90$ \kms.  Considering all the individual \civ\
components observed in nine galaxies, we measure a median component line
width of ${\rm FWHM}(\civ)=49$ \kms.  We note that the observed median
line widths are significantly larger than the line width expected of
thermal broadening in a warm gas of $T\sim 4\times 10^4$ K, which is
$\approx 7.5$ \kms\ for \civ\ absorbing gas.  The observed line widths
are therefore likely driven by the gas dynamics in galactic halos.

We present in Figure 6 the relative velocity distribution of \lya\ and
\civ\ absorbers with respect to the systemic redshifts of the
galaxies.  Here we also adopt the velocity centroid of the strongest
component found in each absorber as the systemic velocity of the
absorber.  The left panel of Figure 6 shows that the velocity
distribution of \lya\ absorbing gas around galaxies is best
represented by a double Gaussian profile with a narrow component
centered at $\langle\,v({\rm Ly\alpha})\,\rangle = -7 $ \kms\ and
dispersion of $\sigma_v({\rm Ly\alpha}) = 32$ \kms\ and a broad
component centered at $\langle\,v({\rm Ly\alpha})\,\rangle = -8 $
\kms\ and dispersion of $\sigma_v({\rm Ly\alpha}) = 188$ \kms\ (the
solid curve). Excluding the outlier at $v({\rm CIV})\approx +260$
\kms, the velocity distribution of \civ\ absorbing gas around galaxies
can be characterized by a single Gaussian distribution of
$\langle\,v({\rm CIV})\,\rangle = 9$ \kms\ and $\sigma_v({\rm CIV})
= 58$ \kms\ (right panel of Figure 6).  

However, we note that the sample size of \civ\ absorbing galaxies is
small.  It is possible that the difference seen between \lya\ and
\civ\ absorbing gas in Figure 6 is driven by sampling uncertainties.
To test this possibility, we randomly generated a large number of
subsamples of nine \lya\ absorbers from the parent sample of 125
\lya\ absorbers, and performed a Kolmogorov--Smirnov (KS) test to
compare each subsample of \lya\ absorbers with the full sample of
\civ\ absorbers.  Roughly 1/3 of the time, the KS test gives $> 95$\%
chance that the two samples are drawn from the same underlying parent
population.  Whether or not \lya\ and \civ\ absorbers share the same
origin remains to be tested with a larger sample of \civ\ absorbing
galaxies.

Considering the \lya\ distribution alone, we find that that the broad
component is dominated by galaxies at $d > 250$ kpc (red open
histogram in the left panel Figure 6). We therefore hypothosize two
different origins of the detected \lya\ absorbers: (1) halo gas
associated with the galaxies contributing primarily to the narrow
Gaussian distribution component and (2) intergalactic gas associated
with correlated large-scale filaments dominating the broad wings.  We
suggest that if larger samples of CIV absorbers were found to lack a
broad velocity component, that would further support this scenario.


We also repeat the search for C\,II\,$\lambda\,1334$,
Si\,II\,$\lambda\,1260$, Si\,III\,$\lambda\,1206$, and
Si\,IV\,$\lambda\,1393$ absorption features associated with these
galaxies.  The observed median line width for C\,II absorbing gas is
${\rm FWHM}\approx 67$ \kms, while the observed median line widths for
silicon ions are ${\rm FWHM}\approx 27, 36$, and 47 \kms\ for Si\,II,
Si\,III, and Si\,IV, respectively.  We therefore adopt these velocity
intervals for measuring 2-$\sigma$ upper limits for respective ionic
transitions\footnote{The velocity window adopted here for measuring
  upper limits of non-detections is empirically determined based on
  detected components, and is on average $4-6$ times smaller than the
  velocity window adopted by Werk \etal\ (2013) for determining their
  upper limits.  As a result, the reported upper limits in Werk
  \etal\ (2013) are roughly 3 times higher than what we determine
  here.}.  The observed absorption properties of the gaseous halos
around individual galaxies are summarized in Table 4, which we list
for each galaxy the angular separation $\theta$ from the QSO
sightline, the projected proper distance $d$, redshift $z_{\rm gal}$,
the redshift of the associated \lya\ absorber if detected, the
measured rest-frame absorption equivalent widths or 2-$\sigma$ upper
limits of different transitions.

\subsection{Absorption Properties of Individual Galactic Halos}

To examine the relation between the incidence and strength of CGM
absorption features and galaxy properties, we present in Figure 7 the
observed distribution of rest-frame absorption equivalent widths of
\lya, C\,II, and \civ\ versus projected distance of the associated
galaxy for all galaxies in our galaxy sample.  The results can be
summarized as follows.

First, robust constraints in \lya\ absorption can be determined for
168 galaxies (solid points in the upper-left panel of Figure 7), 43 of
which do not have a detectable \lya\ absorption feature to a sensitive
upper limit (grey, solid points with arrows).  All 12 galaxies at
$d<100$ kpc exhibit a corresponding \lya\ absorber, leading to a 100\%
mean covering fraction of H\,I absorbing gas within 100 kpc of
star-forming regions.  Beyond $d=100$ kpc, the incidence of \lya\
absorption remains high with only 43 of 156 galaxies showing no
corresponding \lya\ absorption features of $W_r(1215)\apg 0.1$ \AA,
leading to a mean gas covering fraction of $>70$\% over the projected
distance interval of $100-500$ kpc from galaxies.  In addition, the
observed mean \lya\ absorption strength gradually declines with
increasing $d$, consistent with the mean relation reported by Chen
\etal\ (1998, 2001a; solid line).  For comparisons, we include
measurements and upper limits for $\sim\,L_*$ galaxies at
$z=0.14-0.36$ from the COS-Halos program (Werk \etal\ 2013; open
squares) and for starburst galaxies at $z=2.2$ from Steidel \etal\
(2010; green stars).

Second, robust constraints in C\,II absorption can be determined for
141 galaxies (solid points in the middle-left panel of Figure 7), 134
of which do not have a detectable C\,II absorption feature to a
sensitive upper limit (grey, solid points with arrows).  Three of the
four galaxies at $d\apl 60$ kpc exhibit associated C\,II absorption
and none of the eight galaxies at $d=60-100$ kpc show a corresponding
C\,II absorber of $W_r(1334)>0.04$ \AA.  This leads to a mean covering
fraction of $\approx 67$\% in C\,II absorbing gas at $d\apl 60$ kpc of
star-forming regions and 25\% at $d< 100$ kpc.  Beyond $d=100$ kpc, we
detect C\,II absorption near four additional galaxies at $d=106, 116,
138$, and 225 kpc, and the remaining galaxies do not show associated
C\,II absorber to 2-$\sigma$ upper limits of better than $W_r(1334) >
0.1$ \AA.  The lack of observed C\,II absorption leads to a strong
constraint on the mean C\,II absorbing-gas covering fraction of
$\approx 3$\% over the projected distance interval of $100-500$ kpc
from galaxies.  Likewise, we include measurements and upper limits for
$\sim\,L_*$ galaxies at $z=0.14-0.36$ from the COS-Halos program (Werk
\etal\ 2013; open squares) and for starburst galaxies at $z=2.2$ from
Steidel \etal\ (2010; star symbols) for comparisons.

A possible explanation for a substantially lower incidence of C\,II
absorption at $d > 100$ kpc is an increased ionizing radiation field
that offsets a larger fraction of carbon atoms into higher ionization
states.  We examine the incidence of \civ\ absorption features in the
bottom-left panel of Figure 7, where we show that robust constraints
in \civ\ absorption can be determined for 160 galaxies (solid points)
and 151 of these do not have a detectable C\,IV absorption feature to
a sensitive upper limit (grey, solid points with arrows).  Of the 12
galaxies at $d < 100$ kpc, four show a corresponding \civ\ absorber,
leading to a mean covering fraction of $\approx 33$\% in \civ\
absorbing gas within 100 kpc of star-forming regions.  Beyond $d=100$
kpc, five galaxies shows associated \civ\ absorption and the rest do
not show corresponding \civ\ absorption of $W_r(1548)\apg 0.1$ \AA,
leading to a mean \civ\ absorbing-gas covering fraction of $\approx
3.4$\% over the projected distance interval of $100-500$ kpc from
galaxies.  For comparisons, we include measurements and upper limits
for $\sim\,L_*$ galaxies at $z\approx 0.4$ from (Chen \etal\ 2001b;
open triangles) and for starburst galaxies at $z=2.2$ from Steidel
\etal\ (2010; pluses).  Note that the C\,IV doublets were not resolved
in the stacked spectra of Steidel \etal\ (2010).  We therefore display
a possible range of \ewr\ based on two extremes in the assumed line
ratio (e.g. 2:1 or 1:1) between the C\,IV\,$\lambda\,1548$ and
C\,IV\,$\lambda\,1550$ members.

\begin{figure*}
	\begin{center}
%
%
	\includegraphics[scale=0.75]{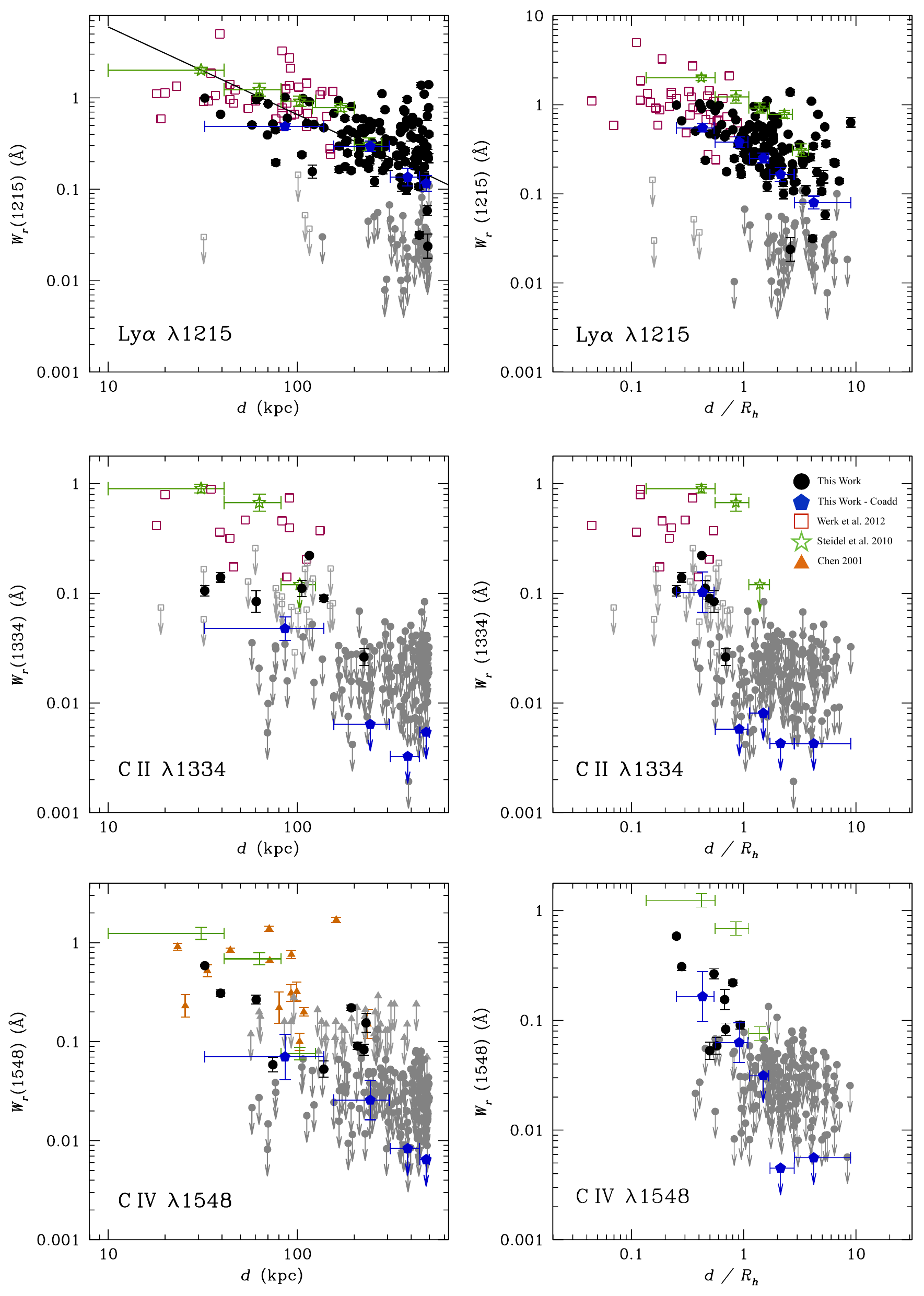}	
	\end{center}
	\caption{Observed spatial distribution of rest-frame
          absorption equivalent widths of \lya\ (top), C\,II (middle),
          and C\,IV\ (bottom) in the circumgalactic space.  The left
          panels show the variation of absorber strengths versus
          projected distance $d$, while the right panels show the
          spatial variation as a function of halo-radius ($R_h$)
          normalized projected distance, $d/R_h$.  Galaxies with no
          detectable absorption are shown as 2-$\sigma$ upper limits
          (points with downward arrows) that are determined based on
          the noise characteristics (see \S\ 4.1).  Measurements from
          our study are shown in solid points.  Measurements for
          $\sim\,L_*$ galaxies at $z=0.14-0.36$ from the COS-Halos
          program (Werk \etal\ 2013) are shown in open squares.  For
          starburst galaxies at $z=2.2$ (Steidel \etal\ 2010),
          constraints on \lya\ and C\,II absorption are shown in star
          symbols.  But because C\,IV doublets were not resolved in
          this study, a range of \ewr\ is shown in vertical bars based
          on assumed doublet ratios of 2:1 and 1:1. The measurements
          at $z=2.2$ are from stacked spectra of galaxies covering a
          range in stellar mass (Reddy \etal\ 2012).  Previous
          measurements of \civ\ absorption in the low-redshift CGM
          from Chen \etal\ (2001b) are shown in triangles.  The
          best-fit power-law model of Chen \etal\ (1998, 2001a) to
          describe the observed $W_r(1215)$ versus $d$ anti-correlation
          is shown in solid line.  Finally, mean absorption strengths
          determined from stacked CGM spectra of the full sample (\S\
          4.4) are shown in (blue) pentagons with the horizontal bars
          indicating the bin size. }
\end{figure*}

\begin{figure*}
	\begin{center}
	\includegraphics[scale=0.78]{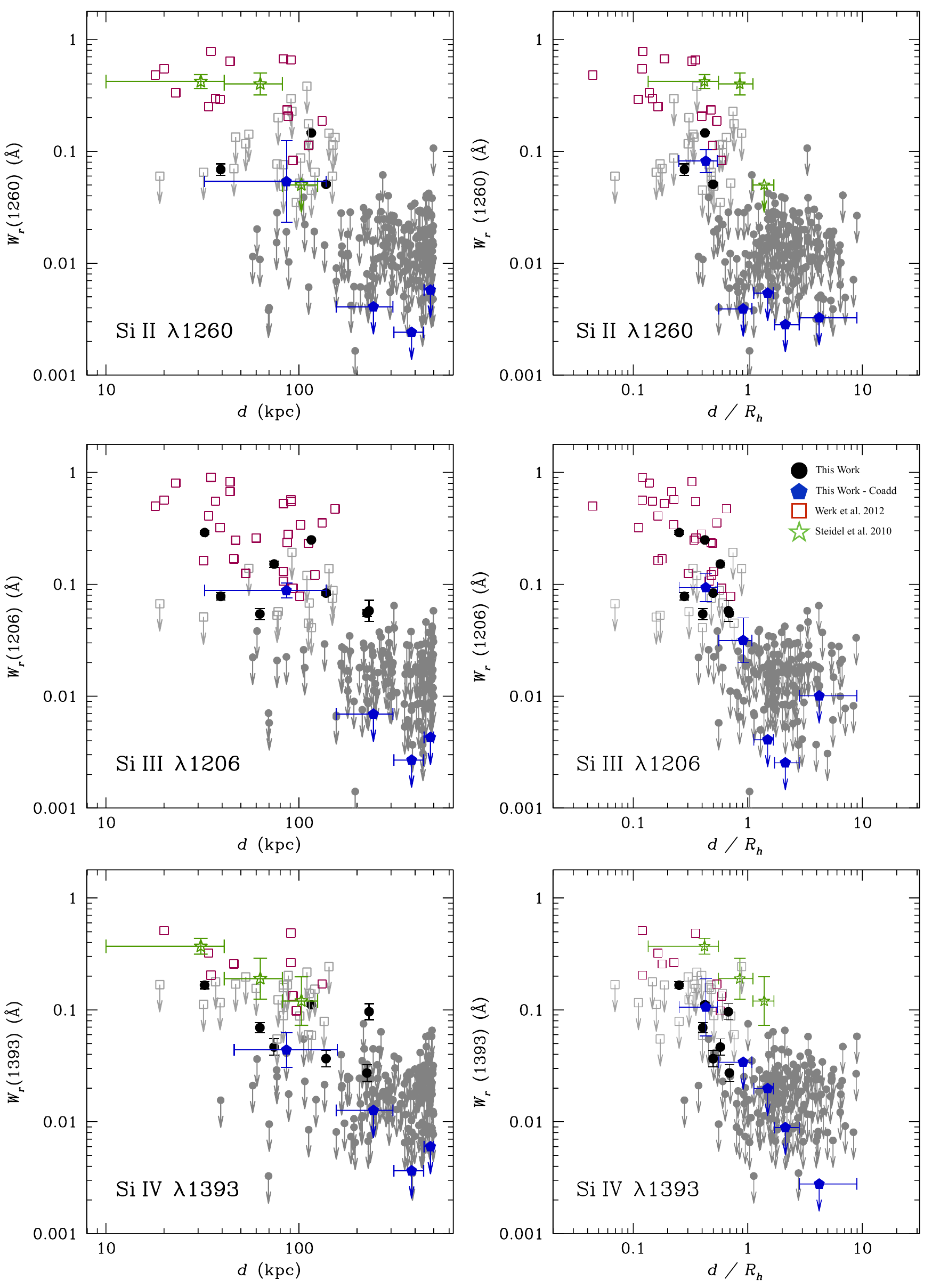}	
	\end{center}
	
	\caption{Similar to Figure 6 but for the observed spatial
          distribution of rest-frame absorption equivalent widths of
          Si\,II (top), Si\,III (middle), and Si\,IV\ (bottom) in the
          circumgalactic space.  Note that the constraints for
          non-detections in the COS-Halos sample appear to be worse
          than our typical upper limits, because the COS-Halos team
          adopted a much larger velocity window for measuring their
          upper limits (i.e.\ $\approx 300$ \kms\ adopted for the
          COS-Halos galaxies versus $\approx 50$ \kms\ adopted in
          our study.  The velocity window adopted in our study is
          empirically determined based on detected components, and is
          therefore more appropriate for constraining the
          presence/absence of cool, photo-ionized clouds in the
          circumgalactic space.}
\end{figure*}

The QSO spectra also allow us to examine the spatial distribution of
silicon ions in the CGM.  Figure 8 displays the observed absorption
strength of Si\,II\,$\lambda\,1260$, Si\,III\,$\lambda\,1206$, and
Si\,IV\,$\lambda\,1393$ versus projected distance of the associated
galaxy for all galaxies in our sample.  We find that the
Si\,II\,$\lambda\,1260$ transition appears to be contaminated for 37
galaxies, and that only three of the remaining 158 galaxies shows
associated Si\,II absorption at $d=39, 115$, and 138 kpc. (upper-left
panel of Figure 8).  With the exception of one galaxy at $d=497$ kpc,
for which no sensitive limit can be placed due to the poor quality of
the data, none of the other 154 galaxies show associated Si\,II
absorption to sensitive limits better than 0.06 \AA.  In addition, the
Si\,III\,$\lambda\,1206$ transition appears to be contaminated for 62
galaxies, and eight of the remaining 135 galaxies show associated
Si\,III absorption.  Of the 10 galaxies at $d<100$ kpc for which
constraints for Si\,III absorption can be obtained, four exhibit
associated Si\,III absorption, leading to a mean Si\,III covering
fraction of $\approx 40$\% (middle-left panel of Figure 8).  At
$d>100$ kpc, four of the 125 galaxies exhibit associated Si\,III,
leading to a mean covering fraction of $\approx 3$\%.  Finally, robust
constraints in the Si\,IV\,$\lambda\,1393$ absorption can be
determined for 150 galaxies, seven of these exhibit associated Si\,IV
absorption with four of the detections found at $d > 100$ kpc
(bottom-left panel of Figure 8).  None of the other 143 galaxies show
associated Si\,IV absorption to sensitive limits better than 0.07 \AA.

The left panels of Figures 7 \& 8 display a stark contrast between the
spatial distributions of hydrogen atoms and heavy elements.  While
hydrogen gas is observed all the way out to 500 kpc in projected
distance, the same galaxies that display moderately strong H\,I
absorption do not exhibit associated heavy elements at $d > 200$
kpc.  
The lack of heavy elements at large distances persists through all
ionization states examined in our study, from low-ionization species
such as C$^{+}$ and Si$^+$ ions to high-ionization species such as
C$^{3+}$ and Si$^{3+}$ ions.  It indicates that if heavy ions are
present in the outer halos, then they remain in a tenuous, hot phase
in which gaseous clumps giving rise to the observed absorption lines
cannot survive.  Similar to Figure 7, we include previous measurements
and upper limits for $\sim\,L_*$ galaxies at $z=0.14-0.36$ from the
COS-Halos program (Werk \etal\ 2013; open squares) and for starburst
galaxies at $z=2.2$ from Steidel \etal\ (2010; star symbols) for
comparisons.

\subsection{Mass Scaling Of Extended Gaseous Halos}

In addition to the distinct spatial extent between hydrogen gas and
heavy ions, the left panels of Figure 7 \& 8 also display significant
scatter in the $W_r$ versus $d$ space.  In particular, different
galaxy samples appear to segregate in different parts of the $W_r$-$d$
parameter space.  Both the $z=2.2$ starburst galaxy sample and the
COS-Halos sample contain primarily galaxies at $d<100$ kpc, which
exhibit on average stronger absorption in \lya\ and in ionic
transitions.  Such distinction appears particularly strong in C\,II
and Si\,II absorption strengths.

Recall that our galaxies span a broad range of stellar mass with 50\%
of the sample characterized as low-mass dwarf galaxies of $M_{\rm
  star}< 0.2\,M_{\rm star}^*$, while the COS-Halos galaxies are
primarily $M_{\rm star}^*$ galaxies (Werk \etal\ 2013) and the $z=2.2$
starburst sample from Steidel \etal\ (2010) has a median stellar mass
that lies between the COS-Halos and our samples (e.g.\ Reddy \etal\
2012; see also Figure 3).  It has been shown empirically at low
redshifts that more massive galaxies on average possess more extended
gaseous halos (e.g.\ Chen \etal\ 2010b).  This is also expected by
theoretical models (e.g.\ Mo \& Miralda-Escud\'e 1996; Maller \&
Bullock 2004; Ford \etal\ 2013b).  As a first step toward establishing
a clear understanding of the processes that drive the observed CGM
properties, it is therefore necessary to investigate whether the
apparent distinction in the observed gaseous extent of different
galaxy samples can be explained by the intrinsic size differences of
individual galactic halos.  The results will facilitate future studies
that consider the effect of additional processes such as star
formation feedback.

To address the possible mass-scaling of gaseous radius, we first
estimate the dark matter halo mass $M_h$ for each galaxy based on its
known stellar mass $M_{\rm star}$ from the NASA-Sloan Atlas and the
stellar mass-halo mass relation of Kravtsov \etal\ (2014) at $z\apl
0.1$.  We then compute the halo radius $R_h$, following the
prescription of Bryan \& Norman (1998)
\begin{equation} 
M_h = \frac{4\pi}{3} \Delta_h\,\langle\rho_{\rm m}\rangle (z) R_h^3,
\end{equation}
where $\langle\rho_{\rm m}\rangle (z)$ is the mean matter density of
the Universe at $z$, $\Delta_h$ represents the overdensity over
which a halo is defined, and
$\Delta_h=18\,\pi^2+82\,x-39\,x^2$ with $x\equiv\Omega_M\,(1+z)^3\,[H_0/H(z)]^2-1$.
The halo radius is then computed according to
\begin{equation} 
R_h = \frac{261.3}{(1+z)}\,\left(\frac{\Delta_h(z)\Omega_M}{97.2}\right)^{-1/3}\left(\frac{M_h}{10^{12}\,{\rm M}_\odot} \right)^{1/3} \rm{kpc}.
\end{equation}
Lastly, we divide the projected distance of each galaxy by its halo
radius and examine how the observed absorption strengths of different
transitions vary with $R_h$-normalized projected distance, $d/R_h$.
The results are shown in the right panels of Figures 7 \& 8.

\begin{figure*}
	\centering
	\includegraphics[scale = 0.9]{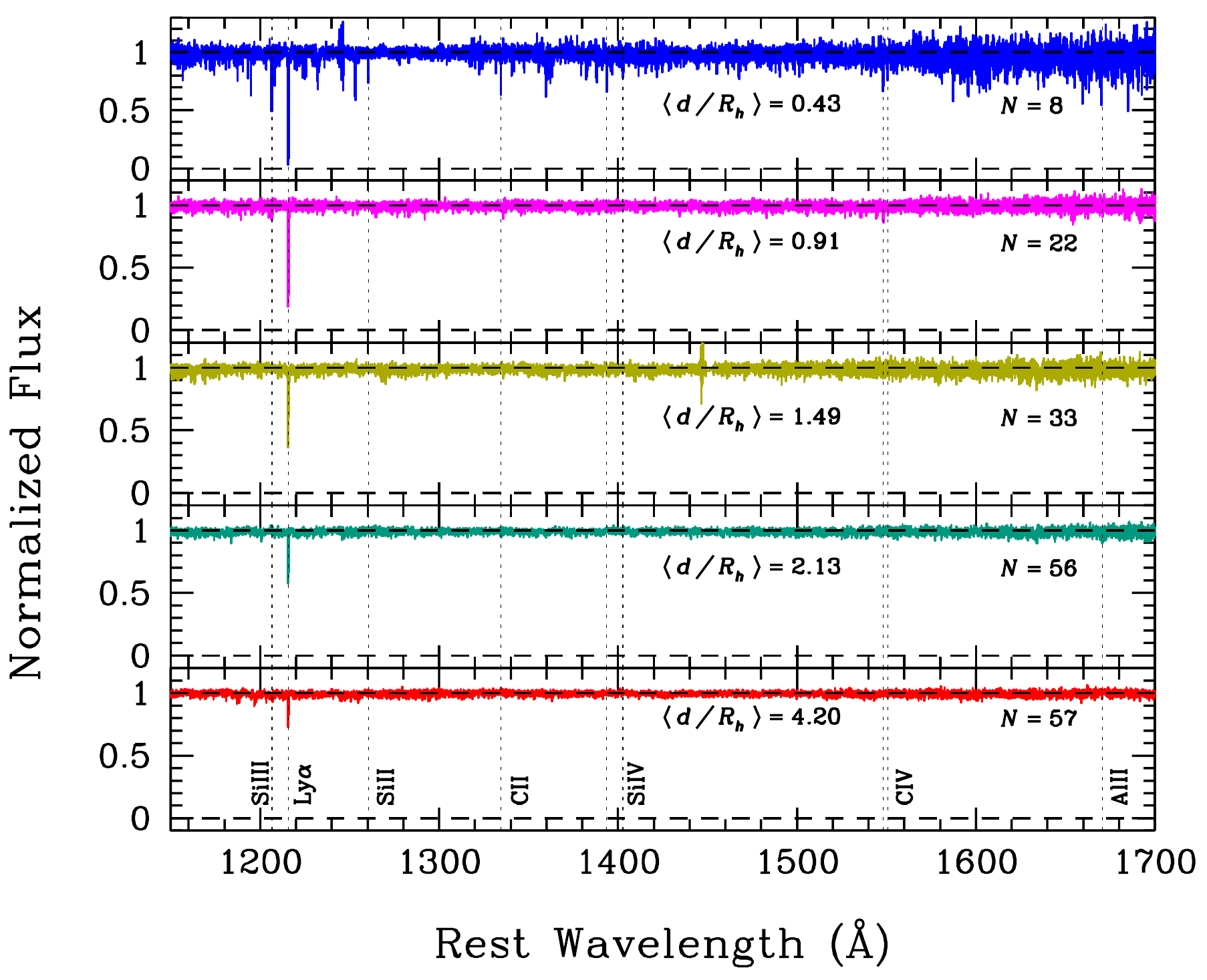}
	\caption{Stacked QSOs spectra in different $R_h$-normalized
          projected distance intervals from host galaxies.  Galaxies
          at $d<1.7\,R_h$ are grouped into three bins following the
          binning of the high-redshift starburst sample.  Galaxies at
          $d>1.7\,R_h$ are divided into two $d/R_h$ bins with roughly
          equal number of galaxies.  The median $R_h$-normalized
          projected distance and the number of galaxies included in
          each stack are indicated in each panel.  The stacked spectra
          clearly show declining \lya\ absorption strengths with
          increasing projected distance, and no ionic transitions are
          detected with $d /R_h > 0.7$.}
\end{figure*}

For comparison, we also calculate the halo radius for each COS-Halos
galaxy using the published $M_{\rm star}$ from Werk \etal\ (2013),
normalize the galaxy projected distance by the corresponding $R_h$,
and include their absorption measurements in the $R_h$-normalized
panels.  In addition, we adopt the median halo mass of
$\langle\log\,(M_{\rm star}\,/\,M_\odot)\rangle=9.9 \pm 0.5$ for the
starburst galaxies at $z=2.2$ from Reddy \etal\ (2012).  Using a
stellar mass-halo mass relation appropriate for high-redshift galaxies
from Behroozi \etal\ (2013), we estimate a mean halo mass of
$\log\,(M_h\,/\,{\rm M}_\odot)=11.6\pm 0.2$ and a mean halo radius of
$R_h \approx 74 \pm 13$ kpc\footnote{The inferred halo mass is a
  factor of two less than what is estimated by Steidel et al.\ (2010)
  or by Trainor \& Steidel (2012) based on the observed galaxy
  clustering amplitude.  For consistency, we adopt the halo mass
  estimate based on the stellar mass to halo mass relation.}.  Because
the dispersion in $R_h$ as a result of dispersion in $M_{\rm star}$ of
the starburst sample is much less that the bin size adopted for $d$,
we normalize their mean projected distance by a single mean halo
radius and include their measurements in the right panels of Figures 7
\& 8.  We cannot include the earlier \civ\ measurements of Chen \etal\
(2001b) because the mass of each galaxy is unknown.

After accounting for a possible mass scaling of gaseous radius, we
note three important features in the right panels of Figures 7 \& 8.
First, the QSOs in our pair sample probe regions as close as
$0.2\,R_h$ in galactic halos to $\approx 10\,R_h$ outside of galactic
halos.  While \lya\ absorbers are frequently observed at projected
distances much beyond $R_h$, no ionic transitions are found at these
large radii.  The few detections at $d>100$ kpc turn out to be
associated with most massive galaxies in the sample and the QSO
sightlines appear to still probe the inner halos of these galaxies.
The large sample allows us for the first time to place stringent
limits on the possible presence of heavy elements beyond galactic
halos.  We place a strict {\it upper limit} of 3\% (at a 95\%
confidence level) on the incidence of $W_r>0.05$ \AA\ metal-line
absorbers over the range of projected distances from $R_h$ to
$9\,R_h$ (see \S\ 4.5 and Figure 11 below).

Second, the COS-Halos sample probes the metal-enriched CGM over the
projected distance range from $\approx 0.05\,R_h$ to $\approx
0.5\,R_h$.  Recall that the median mass of the COS-Halos sample at
$z=0.1-0.4$ is $4\times 10^{10}\,{\rm M}_\odot$ (Werk \etal\ 2013),
while our galaxy sample spans a broad range in stellar mass with a
median $\langle\,M_{\rm star}\,\rangle=5 \times 10^{9}\,{\rm
  M}_\odot$.  Therefore, the COS-Halos sample probes primarily the
inner halos of massive galaxies and our galaxy sample extends to outer
halos of lower- and high-mass galaxies.  The observed absorption
strengths versus $R_h$-normalized projected distance from the
COS-Halos observations now connect smoothly with the measurements for
our sample.  These include H\,I, C\,II, Si\,II, Si\,III, and Si\,IV.
The on average stronger absorbers found around COS-Halos galaxies can
therefore be explained by higher gas density in the inner halos, at
least around massive galaxies.  Considering our sample with the
COS-Halos sample together shows that both the absorption strengths and
incidence (covering fraction) of heavy elements decline steeply beyond
$0.3\,R_h$ and no heavy ions are detected beyond $\approx 0.7\,R_h$.

Finally, the starburst galaxy sample probes the metal-enriched CGM at
$z=2.2$ over the projected distance range from $\approx 0.1\,R_h$ to
$\approx R_h$.  
The mean absorption strength of the metal-enriched CGM around
starburst galaxies at $z=2.2$ is found to decline rapidly at beyond
$R_h$.  In addition, the observed mean absorber strengths at a fixed
$d/R_h$ remain to be stronger around these starburst galaxies than
what is observed around either low-mass dwarf or high-mass galaxies at
$z\sim 0$.  The differential observed absorption strength versus
$R_h$-normalized projected distance between low- and high-redshift
galaxy samples suggests for the first time that the spatial absorption
profile of the CGM may have evolved since $z=2.2$ (cf.\ Chen 2012).
However, a direct comparison between the low- and high-redshift
measurements is difficult because the high-redshift measurements were
made in stacked spectra which presumably includes detections and
non-detections among galaxies with a broad mass range (Table 3), and
there was no distinction made between independent and group galaxies.
An additional uncertainty also arises due to uncertainties in halo
models (Diemer \etal\ 2013; see further discussions in \S\ 5.3).


\subsection{Mean Absorption Profiles from Stacked CGM Spectra}

\begin{figure*}
	\begin{center}
	\includegraphics[scale=0.45]{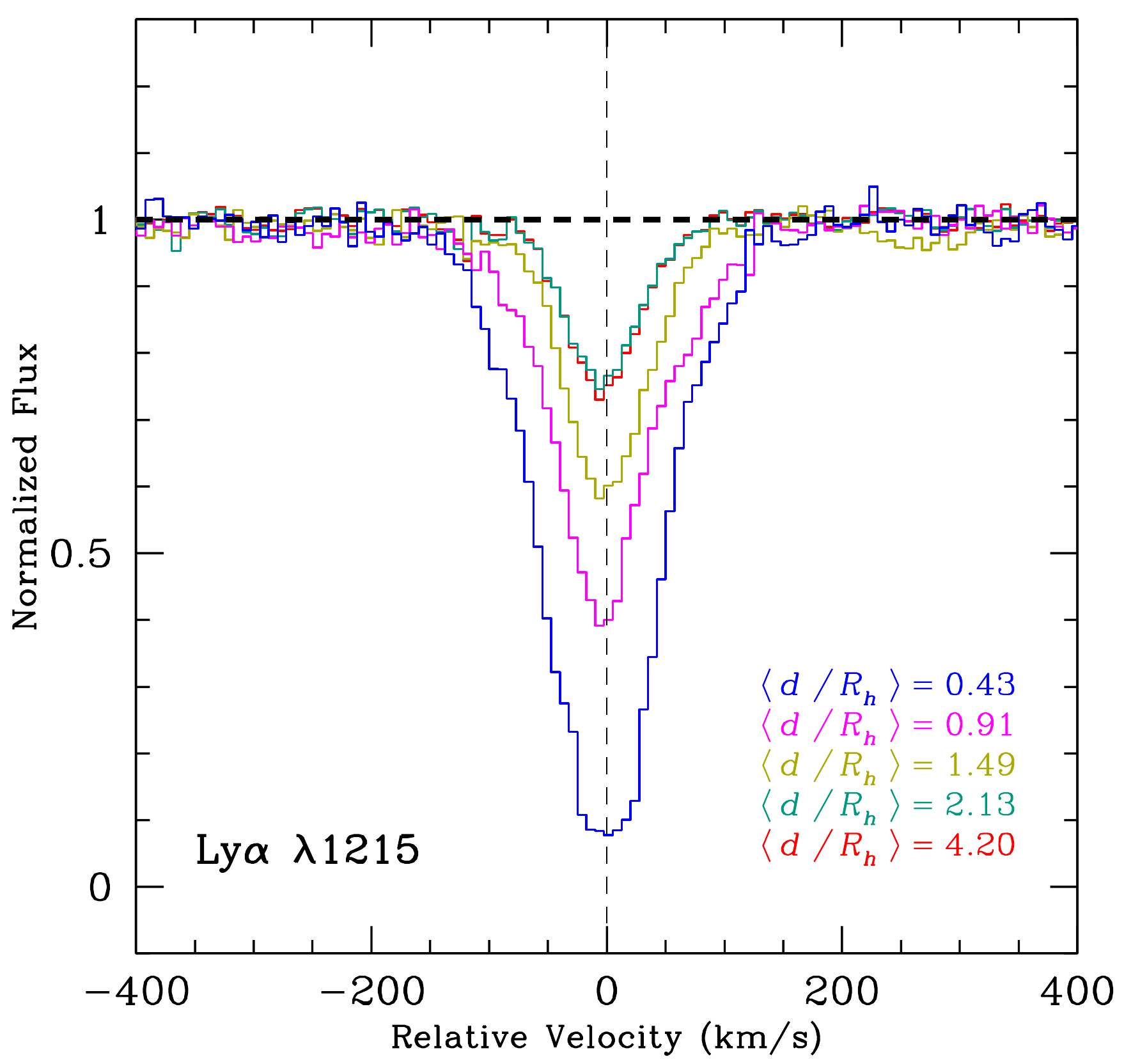}
	\includegraphics[scale=0.45]{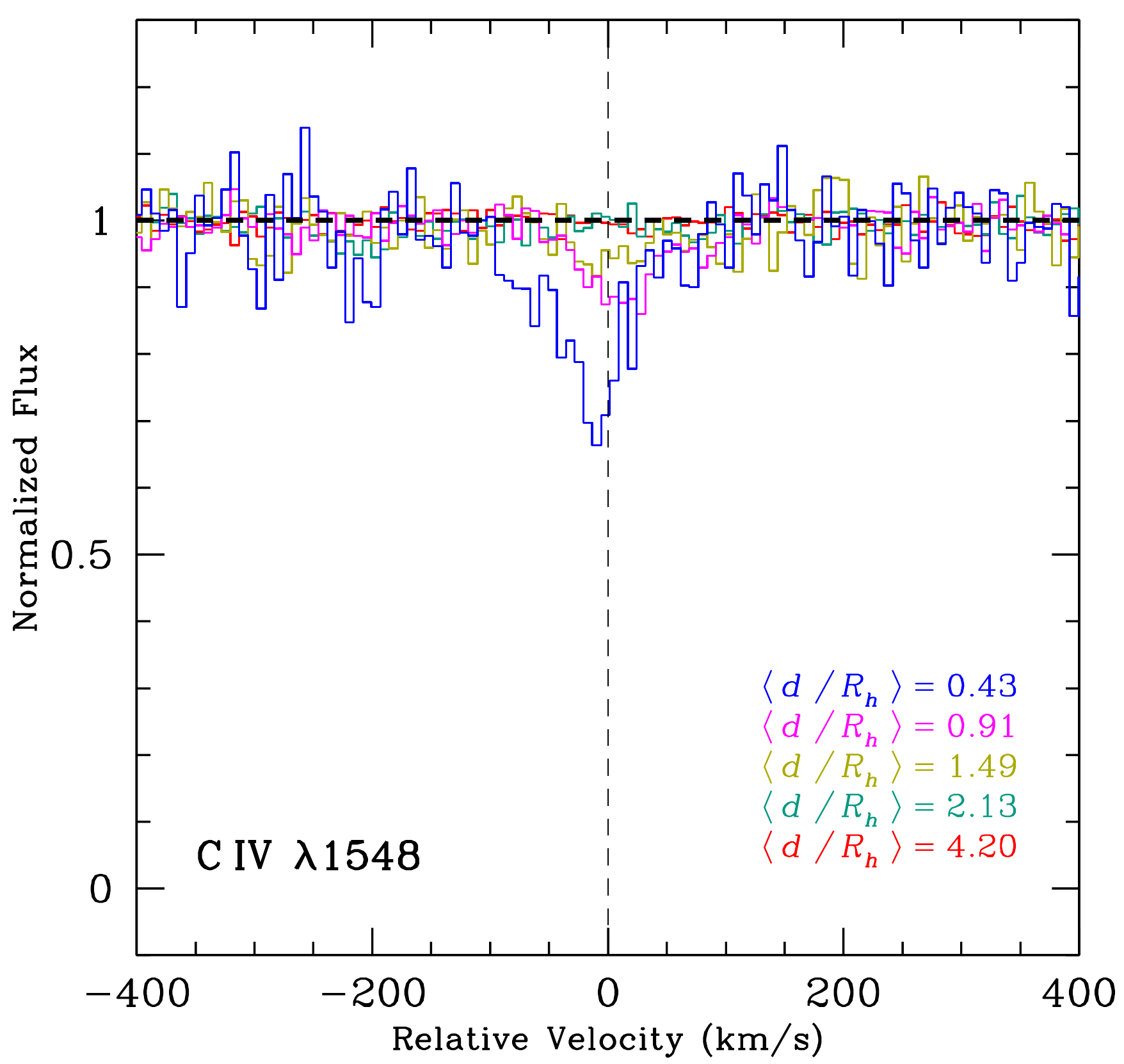}

	\end{center}
	
	\caption{Mean absorption profiles of \lya\ (left) and \civ\
          (right) in stacked absorption spectra of galaxies in
          different $d/R_h$ intervals. }
\end{figure*}

To further improve upon the sensitivity in searching for weak
absorption features and to facilitate a more direct comparison between
our study and that of Steidel \etal\ (2010), we experiment with
co-adding the absorption spectra of individual galaxies in the rest
frame.  We note, however, that interpreting the comparisons between
low- and high-redshift measurements is not trivial (see \S\ 5.3 for
discussion).

For each input spectrum, we first mask contaminating features due to
either the Milky Way interstellar medium or other strong absorbers at
redshifts different from the foreground galaxy.  Then each continuum
normalized QSO spectrum and its associated error spectrum are shifted
to the rest frame of the foreground galaxy.  To account for the
velocity offset between a galaxy and its associated absorbing gas
(Figure 6), the rest frame of each galaxy is set to be the redshift of
the observed \lya\ absorber\footnote{We have also experimented with
  adopting the systemic redshift of the galaxies as the rest frame.
  No significant changes are seen.}.  For galaxies with no detected
\lya\ absorbers, the rest frame is set to be the systemic redshift of
the galaxy.  Individual rest-frame spectra are then weighted and
stacked to form a final mean spectrum.  The weighting factor of each
spectrum is defined as the inverse median variance which is determined
using the associated error spectrum over the wavelength range from
1100 \AA\ to 1450 \AA\ in G130M data and from 1390 \AA\ to 1750 \AA\
in G160M data (e.g.\ columns 5 and 6 in Table 2).

We form different stacked CGM spectra for galaxies grouped in
different projected distance intervals, as well as in different
$R_h$-normalized projected distance intervals.  Given that a large
fraction of detected heavy ions occur at $d<150$ kpc (left panels of
Figures 7 \& 8), we have adopted the first bin in projected distances
to include galaxies at $d<150$ kpc and divided the remaining galaxies
into three projected distance intervals with roughly equal number
($N\approx 50$) of galaxies.  For $R_h$-normalized measurements (right
panels of Figures 7 \& 8), we have adopted the same binning as the
high-redshift starburst galaxy sample at $d<1.7\,R_h$ and divided the
rest of the galaxy sample into two $d/R_h$ bins with roughly equal
number ($N\approx 56$) of galaxies.  The stacked CGM spectra for
different $R_h$-normalized distance intervals are presented in Figure
9.

In each stacked CGM spectrum, we measure mean absorption strengths of
different transitions.  However, uncertainty estimates of the observed
mean absorption strengths in stacked spectra are complicated for two
main reasons.  First, Figures 7 and 8 show that in each $d$ and
$d/R_h$ bin there is a combination of detections and non-detections.
The mean properties of each stacked spectrum are therefore likely
dominated by sampling errors.  In addition, we have adopted an
inverse-variance weighted mean to form each stacked spectrum.  The
final stacks may be dominated by one or two high $S/N$ sightlines.  To
assess these sampling errors, we employ a bootstrap resampling
technique.

For each subsample, we randomly sample the galaxies with replacement
to construct a new subsample that contains the same number of input
galaxies for stacking.  We then measure the mean absorption equivalent
width or a 2-$\sigma$ upper limit if no absorption features are found
in the new stack.  We repeat this exercise $N_{\rm boot}$ times and
examine the distribution of the resulting mean absorption
measurements.  We find that a convergence is reached in the
distribution when $N_{\rm boot}\apg 350$.  We adopt the 1-$\sigma$
dispersion from the bootstrap resampling distribution as the
uncertainties in the observed mean absorption in stacked spectra.  In
the case of non-detections, we quote the 95\% upper bound of the
distribution as the 2-$\sigma$ upper limit to the underlying
absorption strength.  The measurements and associated uncertainty
estimates, as well as upper limits for non-detections are included in
Figures 7 and 8 as solid blue points.  The results are also summarized
in Tables 5 and 6, which record for each distance interval the range
of projected distance $d$ (or $d/R_h$), median projected distance
$\langle\,d\,\rangle$ (or $\langle\,d/R_h\,\rangle$), the number of
galaxies included in the stack, and the mean absorption strengths in
\lya, Si\,III, Si\,II, Si\,IV, C\,II, and C\,IV.  Both Tables 5 \& 6
and Figure 9 confirm that with higher $S/N$ in the stacked spectra the
mean \lya\ absorption strength declines steadily with increasing
projected distance, while no heavy ions are detected at 95\% upper
limits of $W_0\apl 0.03$ \AA\ at $d\apg R_h$.

The strong upper limits presented in Tables 5 and 6 for non-detections
demonstrate that using stacked spectra we are indeed able to improve
the constraints for the presence/absence of absorbing gas by
$\sim\,\sqrt{N_{\rm gal}}$.  No ionic transitions are found outside of
galactic halos to unprecedentedly sensitive limits of better than
$0.03$ \AA\ at the 2-$\sigma$ level of significance.  In addition,
comparing the absorption-line measurements made in stacked CGM spectra
confirms that massive starburst galaxies at $z=2.2$ (green symbols in
the right panels of Figures 7 \& 8) exhibit significantly stronger
mean CGM absorption than the present-day galaxies (solid blue points).
The differences are most pronounced in low-ionization species traced
by C\,II and Si\,II absorption features.  This is in stark contrast to
the comparable mean absorption strengths of Mg\,II absorbers found at
fixed luminosity-normalized projected distances (Chen 2012).  The
discrepancy is less for higher ionization species such as Si\,IV and
C\,IV.  Most interestingly, while we observe strong differential mean
absorption strengths in different ionization states, such as between
Si\,II and Si\,IV and between C\,II and C\,IV, the mean absorption
strengths of different ions appear to be comparable at large distances
from starburst galaxies at $z=2.2$.  The comparable strengths between
low- and high-ionization transitions suggests that either the
ionization condition in the high-redshift CGM is very different or the
lines are saturated (in contrast to the non-saturated lines observed
in our low-redshift sample) and therefore the line widths are driven
by the underlying velocity field of the gas.

\subsection{Gas Covering Fraction}

\begin{figure*}
  \begin{center}
    \includegraphics[scale =0.50]{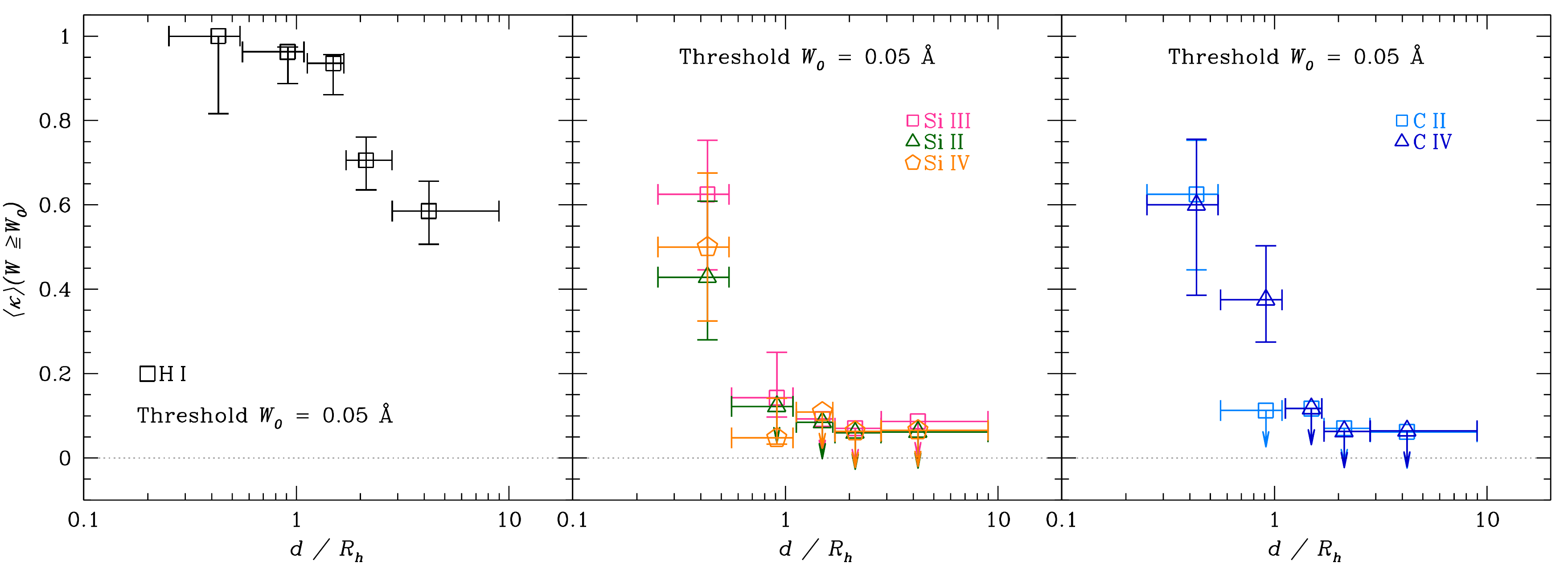}
  \end{center}
  \caption{Mean gas covering fraction \mkap\ measured in each $d/R_h$
    bin for different species.  H\,I is shown in the left panel,
    Si\,II, Si\,III, and Si\,IV are shown in the middle panel, and
    C\,II and C\,IV are shown in the right panel.  Gas covering
    fraction is computed for a detection threshold of $W_0=0.05$ \AA.
    Error bars represent the 68\% confidence interval.  The covering
    fraction of \lya\ absorbing gas remains high at $\approx 60$\%
    outside of dark matter halos at $d>R_h$, while the covering
    fraction of heavy ions is restricted to $\mkap\,<\,10$\% in
    individual $d/R_h$ bins.  Considering {\it all} galaxies at
    $d>R_h$ further improves the limit on the covering fraction of
    heavy ions to $\mkap\,<\,2\%$ at a 95\% confidence level.  In
    inner halos ($d\apl 0.7\,R_h$), we also find a declining mean
    covering fraction from high- to low-ionization species.}

\end{figure*}

We have shown in \S\ 4.4 that stacked CGM spectra enable strong limits
on the possible presence of heavy elements outside of galactic halos,
namely at projected distances $d>R_h$.  Within galactic halos,
however, the observed mean absorption strength in a finite distance
interval is the product of the spatial absorption profile and partial
gas covering.  Figures 7 and 8 show that not only does the absorption
equivalent widths of individual absorbers decrease with increasing
distance, but the fraction of detections also decreases with
increasing distance.  An independent measure of the gas covering
fraction is therefore necessary for an accurate characterization of
the spatial absorption profile of the CGM.

The large number of galaxies in our sample with sensitive limits
available on the presence/absence of extended gas in individual
galactic halos allows us to quantify how the covering fraction
$\kappa$ of absorbing gas varies as a function of projected distance
based on an ensemble average.  Here we consider galaxy subsamples
defined in fixed $d/R_h$ intervals in order to obtain a global average
of how the gas covering fraction depends on halo radius-normalized
distance.

Following the prescription described in Chen \etal\ (2010a), we employ
a maximum likelihood analysis to estimate $\kappa$ for different
species.  First, the probability that a galaxy inside a dark matter
halo of radius $R_h$ exhibits an absorber of rest-frame absorption
equivalent width $W_r\ge W_0$ at projected distance $d$ is written as
\begin{equation} 
P(\kappa|W_0) = \kappa(d/R_h)\,H[r_2-d/R_h]\,H[d/R_h-r_1]
\end{equation}
where $H(x)$ is the Heaviside step function with $H=1$ if $x>0$ and
$H=0$ otherwise, and $r_1$ and $r_2$ together define the distance
interval over which $\kappa$ is calculated.  The likelihood function
of observing an ensemble of $n$ galaxies with associated absorbers of
$W_r\ge W_0$ and $m$ galaxies with no absorbers detected to sensitive
limits of $W_r<W_0$ can then be written as
\begin{eqnarray} 
\mathcal{L}(\kappa|W_0)&=&\Pi_{i=1}^{n}\kappa(d_i/R_h^i)\,H[r_2-d_i/R_h^i]\,H[d_i/R_h^i-r_1] \times \nonumber \\
& &\Pi_{i=1}^{m}\left(1-\kappa(d_i/R_h^i)\,H[r_2-d_i/R_h^i]\,H[d_i/R_h^i-r_1]\right)\nonumber \\
&=&\langle\kappa\rangle^n(1-\langle\kappa\rangle)^m.
\end{eqnarray}

Equation (5) is simply a binomial distribution function of a sample of
$m+n$ objects, in which $n$ are detections, $m$ are non-detections,
and \mkap\ represents the mean probability of detecting an absorber
that maximizes the likelihood $\mathcal{L}$.  In the form of a
likelihood function, we are able to determine the confidence level for
the best fit \mkap.  We evaluate \mkap\ for $W_0 = 0.05$ \AA\ by
maximizing the likelihood function $\mathcal{L}$ for different
absorption transitions.  Figures 7 and 8 show that the majority of the
STIS and COS spectra included in our analysis all have sufficient
$S/N$ to allow strong constraints for ruling out the presence of an
absorber of $W_r>0.05$ \AA\ for all transitions considered in our
study.  In addition, at this sensitive limit of $W_0=0.05$ \AA, we are
already probing tenuous gas of H\,I column density $\log\,N({\rm
  HI})\apl 13$ or chemically enriched CGM of C\,IV column density
$\log\,N({\rm C\,IV})\apl 13.5$, assuming photo-ionized gas of $T\sim
4\times 10^4$ K (e.g.\ Bokensenberg \etal\ 2003).  Figure 11 shows the
estimated \mkap\ as a function of $d/R_h$ for different transitions.
Error bars in \mkap\ represent the 68\% confidence interval.  For
non-detections, we present the 95\% single-sided upper limit.  The
results are also summarized in Table 7.

Figure 11 clearly shows that the covering fraction of \lya\ remains
high ($\approx 60$\%) outside of dark matter halos at $d>R_h$, while
the covering fraction of heavy ions is restricted to $\apl 10$\% in
individual $d/R_h$ bins.  Considering {\it all} galaxies at $d>R_h$
further improves the limit on the covering fraction of heavy ions to
$\mkap_{\rm ions}\,<\,3\%$.  At $d\approx\,0.6\, R_h$, low-ionization
transitions display a rapidly declining covering fraction from
$\mkap\apg 50$\% at $d=0.2-0.6\,R_h$ to $\mkap=6$\% and $<12$\% at
$d=0.6-1.1\,R_h$ for C\,II and Si\,II, respectively.  The high gas
covering fraction at $d\apl 0.6\,R_h$ is consistent with measurements
of $\mkap\,\approx\,58-70$\% for C\,II and Si\,II from the COS-Halos
sample which probes inner galactic halos at $d<0.2\,R_h$, and with
Mg\,II observations at $d\,\apl\,0.3\,R_h$ by Chen \etal\ (2010a).  In
contrast, the covering fraction of C\,IV declines from $\mkap=60$\% at
$d<0.6\,R_h$ to $\mkap=38$\% at $d=0.6-1.1\, R_h$.  {\it Combining
  previous results with our findings suggests that halo gas becomes
  progressively more ionized from $d\apl 0.2\,R_h$ to larger
  distances}.

\section{Discussion}

Based on absorption spectroscopy carried out in the vicinities of 195
galaxies at $z\apl 0.17$, we have obtained strong constraints for the
absorption properties of the CGM at low redshift.  We observe distinct
spatial absorption profiles between H\,I and heavy elements, with
extended \lya\ absorbing gas found at distances far beyond the halo
radius $R_h$ but no heavy ions detected at $d\apg 0.7\,R_h$.  Here we
discuss the implications of our study.

\subsection{Observed Absence of Heavy Elements at $d\apg 0.7\,R_h$
  from Star-forming Regions}

A particularly interesting finding from our absorption-line search is
that while between 60\% and 90\% of the galaxies at $d\apg 0.7\,R_h$
show associated \lya\ absorbers in the QSO spectra, all but two of
these \lya\ absorbers exhibit associated heavy ions to sensitive upper
limits.  The lack of detected heavy ions applies to both
low-ionization species probed by the C\,II and Si\,II absorption
transitions and high-ionization species probed by Si\,III, Si\,IV, and
C\,IV.  It implies that either the chemical enrichment in galactic
halos has a finite edge at $d\approx 0.7\,R_h$ or gaseous clumps
giving rise to the observed absorption lines cannot survive at these
large distances.

In addition, the observed absence of heavy ions at $d\apg 0.7\,R_h$ is found
around galaxies that span over four orders of magnitudes in $M_{\rm
  star}$ (Figure 3).  While the $R_h$-normalized radial profile in
principle accounts for the size-scaling between galaxies of different
mass, it does not include possible variations of halo gas properties
as a result of the star formation history of the galaxies.
Specifically, star formation rate is known to correlate strongly with
galaxy mass (e.g.\ Figure 4) and star formation feedback is expected
to substantially alter the physical properties of halo gas (e.g.\
Agertz \& Kravtsov 2014).  At the same time, high-mass galaxies of
$M_{\rm star}\sim 10^{11}\,M_\odot$ are often found in quiescent mode
with little on-going star formation (e.g.\ Muzzin \etal\ 2013), which
would naturally result in a reduced absorption strength in halos around
these massive galaxies (e.g.\ Gauthier \etal\ 2010).

To investigate possible variation in the $R_h$-normalized radial
absorption profile of the CGM, we divide the galaxy sample into low-
and high-mass subsamples.  Figure 12 shows the \civ\ absorption
properties for low-mass galaxies of $M_{\rm star}\apl 0.1\,M_{\rm
  star}^*$ (filled and open squares) and high-mass galaxies of $M_{\rm
  star}\apg 0.1\,M_{\rm star}^*$ (filled and open circles), where
$M_{\rm star}^*=5\times 10^{10}\,{\rm M}_\odot$ (Baldry \etal\ 2012;
Muzzin \etal\ 2013).  Both low- and hig-mass galaxies exhibit a modest
mean covering fraction of strong ($W_r(1548)>0.1$ \AA) \civ\ absorbers
$\mkap\approx 30-50$\% at $d< 0.7\,R_h$, and no associated \civ\
absorbers at $d>R_h$.  The sample is too small to make a strong
distinction, beyond $R_h$-normalization, in the extent of chemical
enrichment between low- and high-mass halos in our sample.

\begin{figure}
\centering
\includegraphics[scale =0.425]{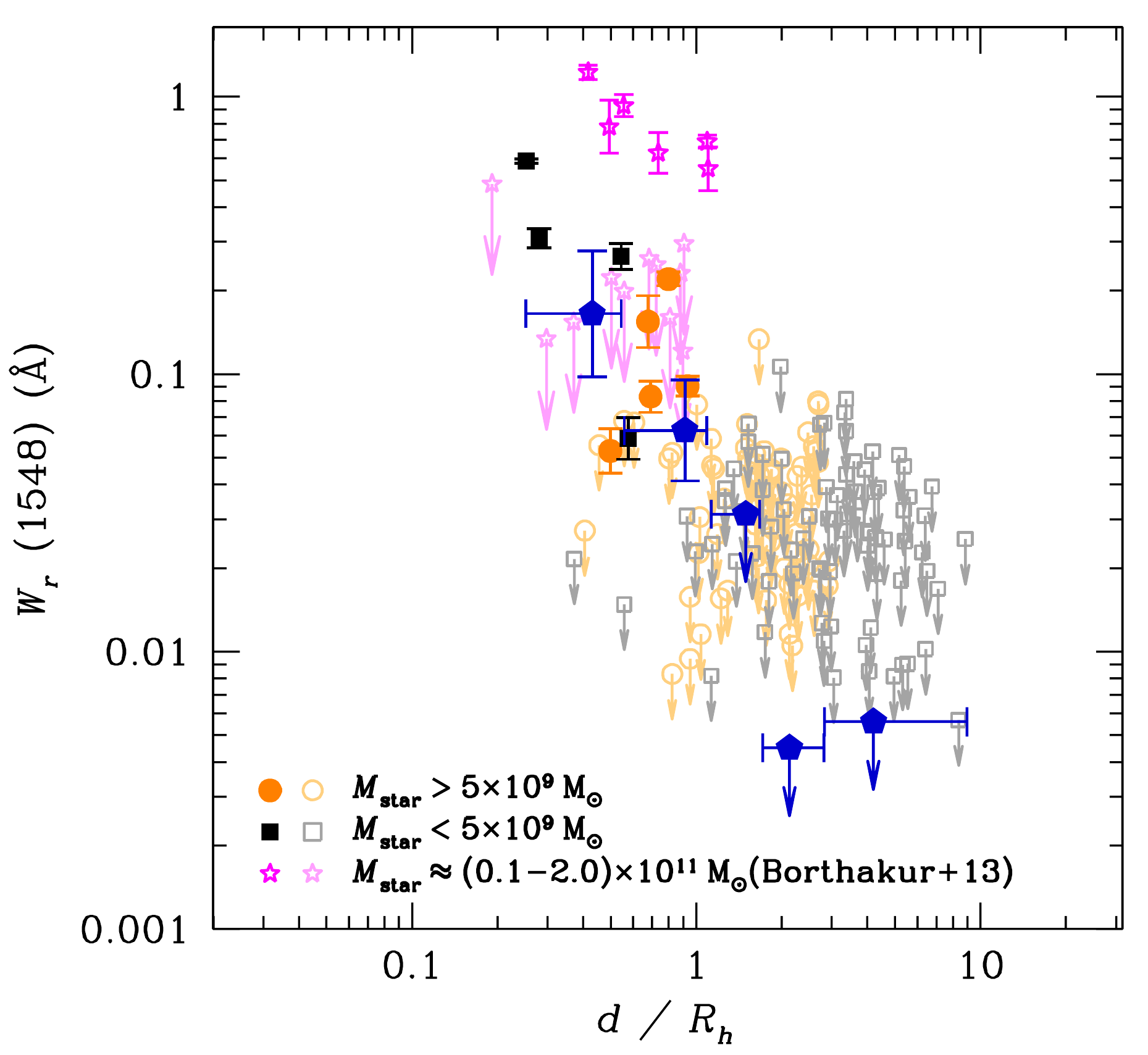}
\caption{$R_h$-normalized radial profiles of \civ\ absorbing gas for
  low-mass galaxies of $M_{\rm star}<0.1\,M_{\rm star}^*$ (filled and
  open squares) and high-mass galaxies of $M_{\rm star}>0.1\,M_{\rm
    star}^*$ (filled and open circles), where $M_{\rm star}^*=5\times
  10^{10}\,{\rm M}_\odot$ (Baldry \etal\ 2012; Muzzin \etal\ 2013).
  Constraints from stacked spectra of the full sample are included for
  reference (pentagon points).  Both low- and hig-mass galaxies
  exhibit a modest mean covering fraction of strong ($W_r(1548)>0.1$
  \AA) \civ\ absorbing gas $\mkap\approx 30-50$\% at $d< 0.7\,R_h$,
  and no associated \civ\ absorbers at $d>R_h$.  No significant
  distinction, beyond $R_h$-normalization, can be made in the extent
  of chemical enrichment around low- and high-mass galaxies.  In
  contrast, six of the 17 galaxies in the Borthakur \etal\ (2013)
  sample exhibit strong \civ\ absorption out to $\sim R_h$ (star
  symbols).  While the constraints for the 11 non-detections are weak,
  the detected \civ\ absorbers around six galaxies (four starburst and
  two passive galaxies) are all significantly stronger than the mean
  values observed in our sample. }
\end{figure}

A lack of heavy ions outside of the halos where dwarf galaxies reside
provides important constraints for the amount of heavy elements that
can escape these low-mass halos.  Starburst driven outflows are
thought to be particularly effective in removing baryons from low-mass
halos because of a shallow potential well (e.g.\ Larson 1974; Martin
1999).  Energetic outflows are also expected to suppress continuing
growth of low-mass galaxies due to a reduced accretion rate
(e.g.\ Scannapieco \etal\ 2001).  These effects together can explain
known scaling relations such as the mass-metallicity relation found in
dwarf galaxies (e.g.\ Dekel \& Woo 2003).  However, model that predict
enrichment levels to $>0.1$ solar metallicity are inconsistent with
the observed limits we obtain at $d>0.7\,R_h$ from low-mass dwarf
galaxies at $z\sim 0$.  This places an important constraint on
feedback models that would enrich the CGM to much higher metallicity
at these distances.

Our finding of a lack of detected heavy elements beyond $d\approx 0.7\,R_h$
appears to be discrepant from the observations of Borthakur \etal\
(2013), who reported detections of strong \civ\ out to $d\sim R_h$
from massive galaxies of $\log\,(M_{\rm star}/M_\odot)\approx 10.1$
(star symbols in Figure 12).  A possible explanation is to attribute
the observed \civ\ absorbers to outflows driven by on-going
star-formation activity, which indeed was part of the selection
criterion of the Borthakur \etal\ sample.  However, two of the
galaxies with associated strong \civ\ absorbers are classified as
quiescent galaxies with little on-going star formation since recent
past.

An alternative scenario different from starburst driven outflows in
explaining the presence of heavy elements at large distances is
through gas stripping in an overdense group environment (e.g.\ Kawata
\& Mulchaey 2008; McCarthy \etal\ 2008).  Such scenario can be tested
with deep galaxy survey data of fields around these strong metal-line
absorbers (e.g.\ Gauthier 2013).

To improve upon the detection limits of heavy ions in \lya\ absorbing
gas, we repeat the stacking exercise discussed in \S\ 4.4 but consider
only galaxies at $d\apg R_h$ which exhibit associated \lya\ absorbers.
The goal of this exercise is to investigate whether some constraints
can be obtained on the chemical enrichment level in known gaseous
clouds revealed by \lya\ absorption.

The absorption-line measurements and constraints in stacked spectra of
\lya\ absorbers are summarized in Table 8.  We divide the galaxies at
$d\apg R_h$ into three subsamples, each covering different
$R_h$-normalized projected distance intervals.  The goal is to
maximize the $S/N$ of the stacked spectra, while searching for
possible trend of chemical enrichment with distance.  In this
particular exercise, we also search for the presence of the
N\,V\,$\lambda\lambda\,1238, 1242$ doublet features.  Similar to the
O\,VI absorption doublet, the N\,V doublet is seen in highly ionized
gas and is covered in the STIS and COS FUV spectral range.
Observations of the N\,V absorption doublet together with C\,IV and
other low-ionization transitions, therefore, provides a measure of the
ionization state of the gas (see for example Figure 32 of Thom \& Chen
2008).  The results of our search remains the same; namely no ionic
transitions are detected in \lya\ absorbers at $d\apg R_h$ from
galaxies in our sample.  Adopting the absorption constraints for \lya\
and C\,IV from Table 8 and assuming typical Doppler parameters,
$b({\rm H\,I})=30$ \kms\ (e.g.\ Tilton \etal\ 2012) and $b({\rm
  C\,IV})=7$ \kms\ (e.g.\ Boksenberg \etal\ 2003), we infer a column
density ratio of $\log\,N({\rm C\,IV})/N({\rm H\,I})<-1.3$.  In
contrast, cosmological simulations that incorporate strong outflows to
reproduce statistical properties of the general galaxy population in
the low-redshift universe predict a column density ratio of
$\log\,N({\rm C\,IV})/N({\rm H\,I})\sim -0.6$ at the virial radius of
galactic halos of $10^{11-12}\,{\rm M}_\odot$ (Ford \etal\ 2013b).
For photo-ionized gas, this column density ratio implies that the
chemical enrichment level of \lya\ absorbing gas at $d\apg R_h$ cannot
exceed 0.1 solar metallicity.

\subsection{The Origin of a Chemical Enrichment Edge at $\sim 0.7\,R_h$}

The distinct boundary between the presence and absence of metal-line
absorbers at $d\approx 0.7\,R_h$ is seen in all ionic transitions that
have been studied, including the observations of Si\,II, Si\,III,
Si\,IV, C\,II, and C\,IV reported here and those of Mg\,II absorbers
around $\langle\,z\,\rangle=0.24$ galaxies (Chen \etal\ 2010a).  It
is, however, not clear whether such absorption boundary exists for
highly ionized gas probed by the O\,VI absorption doublet.  Existing
observations are primarily limited to inner galactic halos at $d\apl
0.5\,R_h$ (e.g.\ Tumlinson \etal\ 2011, 2013).  Other studies that
extend to larger radii have only small samples of ``isolated''
galaxies (e.g.\ Chen \& Mulchaey 2009; Prochaska \etal\ 2011) and
therefore available constraints are uncertain (see Johnson \etal\
2014, in preparation).

A natural explanation for the chemical enrichment edge at $d\approx
0.7\,R_h$ is the turn-around radius $r_{\rm turn}$ of starburst driven
outflows, similar to a halo fountain phenomenon (cf.\ Bregman 1980).
We note that as chemically-enriched outflowing material moves into
low-density regions and mixes in with metal-poor gas, it is expected
that the outflowing material would be diluted and the observed
absorption strength would gradually decline.  However, to produce the
sharp downturn at $d\approx 0.7\,R_h$ would require a sharp decline in
metallicity, which essentally indicates a finite edge of chemical
enrichment and motivates the consideration of a turn-around radius of
starburst driven outflows.

In this scenario, we can infer the launch speed as a function of
launch radius of the outflows in individual halos.  Adopting a
Navarro-Frenk-White (NFW; Navarro \etal\ 1997) density profile for the
underlying dark matter halos, we estimate the launch velocity $v_{\rm
  out}$ at a launch radius $r_i$ following $v_{\rm
  out}(r_i)=\sqrt{2\,[\Phi(r_{\rm turn})-\Phi(r_i)]}$, where $\Phi$ is
the gravitational potential and is related to radius $r$ according to
$\Phi(r)=-(GM_s/r)\times\ln\,(1+r/r_s)$.  For an NFW profile, $r_s$ is
the scale radius and $M_s$ is the total mass enclosed within $r_s$ and
is related to the halo mass according to
$M_h=M_s\,[\ln\,(1+R_h/r_s)-R_h/(r_s+R_h)]$.

For galaxies in halos of $M_h=10^{11}\,{\rm M}_\odot$, which is
typical of the mass range probed by our sample, we calculate a launch
velocity of $v_{\rm out}\approx 170$ \kms\ at $r_i=1$ kpc.  For
lower-mass galaxies in halos of $M_h=10^{10}\,{\rm M}_\odot$, we find
$v_{\rm out}\approx 75$ \kms\ at $r_i=1$ kpc.  For higher-mass
galaxies in halos of $M_h=10^{12}\,{\rm M}_\odot$, we find $v_{\rm
  out}\approx 330$ \kms\ at $r_i=1$ kpc.  Naturally, a larger launch
speed is required in more massive halos in order to reach out to a
larger distance.  Because more massive galaxies on average are forming
stars at a higher rate (e.g.\ Salim \etal\ 2007), it is conceivable to
have a higher outflow speed from starburst driven winds in a more
massive galaxy.  With a larger sample, we will be able to examine
whether/how the metal-line boundary depends on star formation
activity.

An alternative explanation for the observed finite metal-line boundary
at $d\approx 0.7\,R_h$ is the critical radius below which cool clouds
can develop and stabilize in a multi-phase medium (e.g.\ Maller \&
Bullock 2004; McCourt \etal\ 2012).  Specifically, McCourt \etal\
(2012) showed that hot halos can develop extended multiphase
structures if the cooling time $t_{\rm cool}$ is comparable to or less
than the dynamical time $t_{\rm ff}$.  Given that $t_{\rm cool}\propto
1/\rho$ and $t_{\rm ff}\propto 1/\sqrt{\rho}$, the ratio $t_{\rm
  cool}/t_{\rm ff}$ decreases rapidly with decreasing radius.  While
the details depend on the exact density profile of the hot halo, it is
conceivable that such condition can be reached at $\approx 0.7\,R_h$.

Both the turn-around radius of starburst outflows and cooling radius
of a hot halo can provide a physical explanation for the observed
metal-line boundary.  Because the former requires on-going star
formation to provide the energy input, one can in principle
distinguish between the two scenarios by examining whether or not the
presence of a metal-line boundary depends on the star formation
activity in the galaxies.
  
\subsection{The CGM at Low and High Redshifts}


When comparing a luminosity-normalized spatial extent of the CGM at
different redshifts, Chen (2012) found that no distinctions can be
made between low- and high-redshift galaxies that have dissimilar star
formation histories.  The adopted luminosity normalization was meant
to account for a likely luminosity-size scaling relation commonly seen
in optical disks (e.g.\ Cameron \& Driver 2007).  This approach was
motivated by the expectation that the CGM is regulated by various
complicated physical processes, including gravitational motion and
star formation feedback.  Normalizing the observed projected distance
by the intrinsic luminosity offers a means of accounting for the
intrinsic size differences of individual galactic halos, and
represents the first step toward establishing a clear understanding of
the processes that drive the observed CGM properties.  The results
will facilitate future studies that consider the effect of additional
processes such as star formation feedback.

But while rest-frame $B$-band luminosity ($L_B$) is a good proxy of
mass as shown in Chen \etal\ (2010b), it is an indirect measure of
galaxy mass.  Our study presented here utilizes a sample of galaxies
with known stellar masses and normalizes the observed extent of halo
gas by the inferred dark matter halo radius under a simple spherical
overdensity approximation (e.g.\ Maller \& Bullock 2004).

The finding in \S\ 4.3 is surprisingly very different from the finding
of Chen (2012).  Specifically, when accounting for a size-scaling
based on halo mass (instead of rest-frame $B$-band luminosity),
high-redshift starburst galaxies exhibit both systematically stronger
mean absorption strength at a fixed $R_h$-normalized projected
distance at $d/R_h\apl 0.7$ and more extended chemically-enriched gas
to $d\apg R_h$ than low-redshift galaxies.  Furthermore, the
differences in the mean absorption strengths of the low- and
high-redshift CGM are most pronounced in low-ionization species traced
by C\,II and Si\,II absorption lines, suggesting distinct ionization
conditions between the CGM at $z\approx 0$ and at high redshifts near
starburst galaxies.

We note, however, that a fundamental difference between the low- and
high-redshift measurements is the treatment of galaxies with close
neighbors.  In our study presented here, we have considered only
``isolated'' galaxies with no known close neighbors spectroscopically
identified within 500 kpc in projected distance and $|\Delta\,v|<500$
\kms\ in velocity separation.  No such distinction was made in the
high-redshift sample, and therefore the high-redshift CGM measurements
include both isolated galaxies and those with neighbors.  It has been
suggested that galaxies in group environments display a more extended,
chemically enrichment halo gas than galaxies in the field (e.g.\ Chen
\etal\ 2010a; Bordoloi \etal\ 2011).  At the same time, the overall
cool gas content is also found suppressed in galaxy groups (e.g.\
Chynoweth \etal\ 2011).  It is not clear that the more extended,
strong CGM absorption around high-redshift starburst galaxies is due
to a signigicant contribution of galaxies in group environments.  A
detailed study of the CGM in different group environments at low
redshifts will be presented in a separate paper (Liang \etal\ 2014 in
preparation).

In addition, Steidel \etal\ (2010) noted that the Si\,II absorption
lines are saturated at $d\apl 40$ kpc in the stacked low-resolution
spectra.  On the other hand, the ionic transitions observed in
individual galaxies at low redshifts are not commonly saturated.
Steidel \etal\ interpreted the observed declining absorption
equivalent width (at least from $d=0$ to $d=40$ kpc) as likely due to
a combination of decreasing line-of-sight velocity spread and
decreasing covering fraction of the absorbing gas.  In contrast,
observations individual galaxies at low redshift indicate that the
declining mean absorption equivalent width in stacked spectra is best
explained by a declining gas covering fraction along with declining
optical depth with increasing projected distance.  Consequently,
interpreting the observed difference between low- and hig-redshift
observations becomes more complicated.

At $d>40$ kpc, however, Steidel \etal\ (2010) also pointed out that
the ionic absorption lines appeared to become unsaturated around
high-redshift starburst galaxies.  For unsaturated lines in stacked
spectra, the observed declining equivalent width is best explained by
a declining gas covering fraction along with a declining optical
depth, similar to the low-redshift observations.  Comparisons between
our means stacked spectra and those from Steidel \etal\ could
therefore be made for observations at $d>40$ kpc.

Remaining caveats include redshift uncertainties.
Individual spectra of these high-redshift galaxies are of low $S/N$
and low spectral resolution with ${\rm FWHM}\approx 350$ \kms.  Random
motions between galaxies and the absorbing gas (e.g.\ Figure 6) can
smooth out the signal.  Chen (2012) performed a simple Monte Carlo
simulation to show that these uncertainties can result in an overall
reduction of the absorption strength by $\approx 20$\%, increasing the
discrepancy between low- and high-redshift samples.  At the same time,
uncertainties in sky subtraction of low $S/N$ spectra may also bias
the continuum estimate that is difficult to assess.  It is expected
that these uncertainties can be resolved with absorption spectroscopy
of background QSOs near individual starburst galaxies (e.g.\ Rudie
\etal\ 2012; Turner \etal\ 2014).

\begin{figure*}
\centering
\includegraphics[scale =0.45]{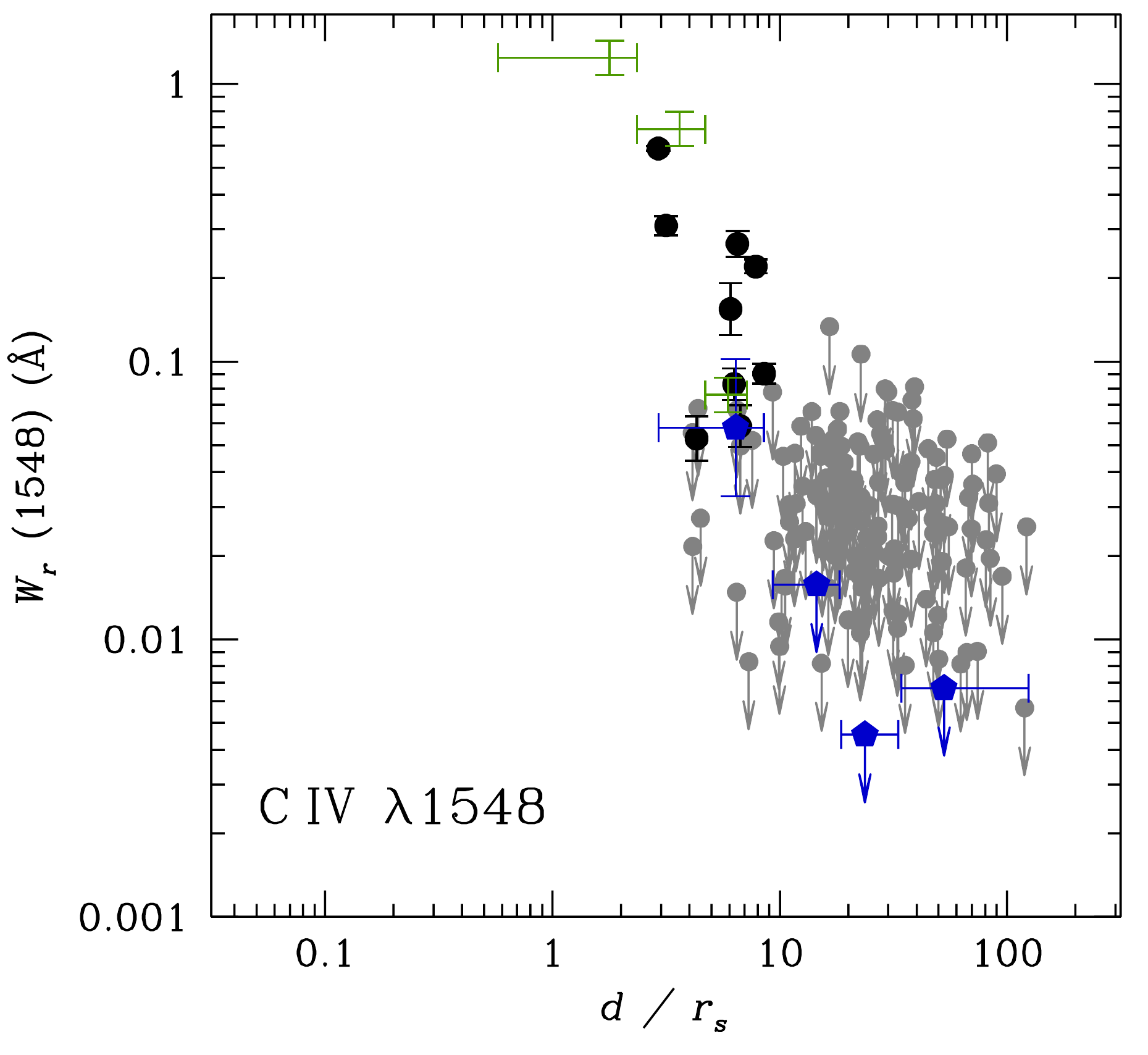}
\includegraphics[scale =0.45]{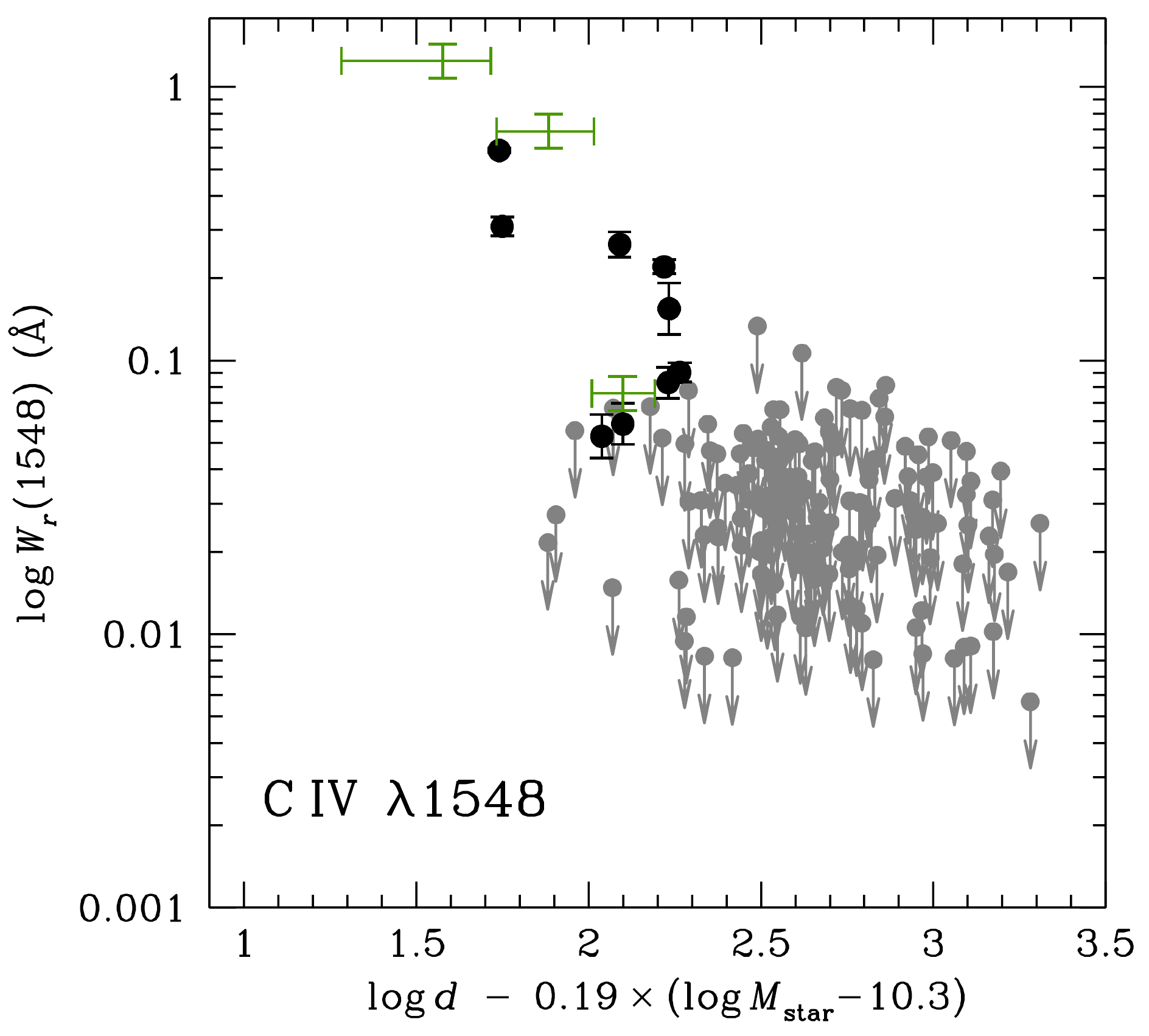}
\caption{\civ\ radial absorption profile normalized by the scale
  radius $r_s$ of the host dark matter halos ({\it left}) or by
  $M_{\rm star}$ of the absorbing galaxies ({\it right}).  The scale
  radius $r_s$ represents a characteristic radius where the dark
  matter density profile resembles an $r^{-2}$ profile (Navarro \etal\
  1997).  We also include in the left panel measurements from stacked
  spectra using the full galaxy sample (blue pentagons).  In addition,
  the scaling coefficient of $M_{\rm star}$ is not a best fit to the
  \civ\ data, but adopted from the best-fit coefficient of Mg\,II
  absorbers from Chen \etal\ (2010b).  Nonetheless, the result already
  reveals a clear \civ\ enriched halo out to $M_{\rm star}$-normalized
  projected distance of $\approx 160$ kpc (consistent with the finding
  of $L_B$-normalized \civ\ profile from Chen 2012).  In both panels,
  the \civ\ radial absorption profiles at $\langle\,z\,\rangle=0.041$
  (solid points) and at $\langle\,z\,\rangle=2.2$ (green crosses)
  appear to be better aligned than what is seen in the bottom right
  panel of Figure 7 \& 8. }
\end{figure*}

\subsection{Uncertainties due to Pseudo-evolution of Halo Radius $R_h$}

We note, however, that a fundamental caveat in comparing
$R_h$-normalized CGM profiles at different redshifts is in how $R_h$
is calculated.  Equations (2) and (3) outlines a commonly adopted
approach to calculate halo mass $M_h$ and halo radius $R_h$ based on a
spherical overdensity model.  A dark matter halo is defined as a
collapsed and virialized object with a mean enclosed density equal to
$\Delta_h$ times the mean background density $\langle\rho_m\rangle$ at
the redshift of interest.  By this definition, however, as the
universe expands, $\langle\rho_m\rangle$ decreases.  Consequently,
both $M_h$ and $R_h$ would increase with time even for a static halo.
Such pseudo-evolution of dark matter halos has been investigated by
several groups, and a consistent finding in numerical simulations is
that most halos of $M_h\apl 10^{12}\,M_\odot$ completed matter
accretion by $z\sim 1$ with little/no net growth in mass over the last
seven billion years (e.g.\ Prada \etal\ 2006; Diemand \etal\ 2007;
Diemer \etal\ 2013).  Therefore, applying an $R_h$ normalization may
impose a pseudo-evolution in the properties of galaxies residing in
these low-mass halos.




Given the uncertainties in $R_h$, we explore different scaling
relations for an accurate comparison of the CGM absorption profiles in
different epochs.  In Figure 13, we present \civ\ radial absorption
profile normalized by the scaled radius $r_s$ of the host dark matter
halos (left panel) or by $M_{\rm star}$ of the absorbing galaxies
(right panel).

The scale radius $r_s$ represents a characteristic radius where the
dark matter density profile resembles an $r^{-2}$ profile (Navarro
\etal\ 1997).  The $r_s$ normalization is motivated by the
understanding that when a halo stops accreting, the scaled radius
$r_s$ remains constant while the halo mass $M_h$ and halo radius $R_h$
continue to increase as the universe expands and
$\langle\rho_m\rangle$ descrases.  As a result, the halo concentration
$c\equiv R_h/r_s$ would also increase.  Indeed, halo concentrations
are found to increase rapidly for halos of $M_h\apl 10^{12}\,M_\odot$
at $z\apl 1$ (e.g.\ Zhao \etal\ 2009).

To calculate $r_s$, we adopt the mean $c$-$M_h$ relation of Zhao et
al.\ (2009), which parameterizes $c$ as a function of halo mass and
redshift, $c(M_h, z)$.  The left panel of Figure 13 shows that the
$r_s$-normalized \civ\ radial profile at $\langle\,z\,\rangle=2.2$
declines rapidly at $d\approx (4-6)\,r_s$.  This is consistent with
observations of the $\langle\,z\,\rangle=0.041$ sample that display a
covering fraction of \civ\ absorbing gas at $d < 6\,r_s$ with
$\mkap_{\rm C\,IV}\approx 50$\% and no detections found beyond
$8\,r_s$.

The agreement in the $r_s$-normalized CGM absorption profiles at low
and high redshifts is in stark contrast to the distinction seen in the
$R_h$-normalized CGM profiles presented in Figures 7 \& 8.  We note,
however, an inherent uncertainty in $r_s$ as a result of the scatter
in the mean $c$-$M_h$ relation ($\sim\,0.15$ dex).  Nonetheless, it
underscores the fundamental difficulty in making direct comparisons of
galaxies and their halo gas properties in two different epochs.

A pseudo-evolution in dark matter halo mass (and therefore halo radius
$R_h$) has important implications beyond a constant CGM radial
absorption profile.  It would also suggest that chemical enrichment
have reached out to the ``edge'' of the dark matter halo ($\sim
4\,r_s$) or likely beyond, although a clear border for the observed
metal-line absorbers outside of halo radius still implies a finite
extent of chemical enrichment around galaxies.  To quantify the extent
of chemical enrichment, we will rely on numerical simulations as a
guide for determining the physical size of the underlying dark matter
halos of our galaxies.  Such study is beyond the scope of this paper.
We defer the analysis to a future paper (Liang et al.\ 2014, in
preparation).

Given the uncertainty in $R_h$ and $r_s$ from dark matter halo models,
we return to applying empirical properties of the galaxies for scaling
halo gas properties.  Instead of $B$-band luminosity, however, we
consider normalizing the projected distance of galaxies by stellar
mass, $M_{\rm star}$.  The right panel of Figure 13 shows the $M_{\rm
  star}$-normalized \civ\ radial profiles for the high- and
low-redshift galaxy samples.  We note that the adopted scaling
coefficient of $M_{\rm star}$ is not a best fit to the \civ\ data, but
adopted from the best-fit coefficient of Mg\,II absorbers from Chen
\etal\ (2010b).  Nonetheless, the result already reveals a clear \civ\
enriched halo out to $M_{\rm star}$-normalized projected distance of
$\approx 160$ kpc (in an excellent agreement with the finding of
$L_B$-normalized \civ\ profile from Chen 2012).  Similar to the
$r_s$-normalized profiles in the left panel, the $M_{\rm
  star}$-normalized \civ\ radial absorption profiles at
$\langle\,z\,\rangle=0.041$ and at $\langle\,z\,\rangle=2.2$ appear to
be well aligned.  With in mind the uncertainty in $R_h$ and the
agreement in $M_{\rm star}$-normalized CGM radial profile, we
therefore conclude that no strong evidence has been found to show a
vastly different spatial distribution in the CGM absorption
properties, despite a completely disparate star formation properties.



\section{Summary}

We have analyzed archival UV spectra of QSOs located at projected
distance $d\le 500$ kpc from 195 independent galaxies in the
foreground.  The redshifts of the galaxies range from $z = 0.002$ to
$z = 0.176$ with a median of $\langle\,z\,\rangle = 0.041 $, and the
projected distances of the QSOs range from $d\approx 32$ kpc to
$d\approx 500$ kpc with a median of $\langle\,d\,\rangle=362$ kpc.
The galaxy sample covers a broad range in stellar mass from $M_{\rm
  star} = 1.5\times 10^5\,{\rm M}_\odot$ to $M_{\rm star} = 1.4 \times
10^{11}\,{\rm M}_\odot$.  The median and 1-$\sigma$ dispersion of
stellar masses in the sample are $\log\,(M_{\rm star}/M_\odot) = 9.7
\pm 1.1$.  The available UV spectra of the QSOs allow us to study the
multiphase CGM based on observations of a suite of absorption features
including \lya, C\,II, C\,IV, Si\,II, Si\,III, and Si\,IV.  Using this
large QSO and galaxy pair sample, we have obtained strong constraints
for the absorption properties of the CGM at $z\apl 0.176$.  The
results of our analysis are summarized as the following:

(1) We observe a stark contrast between the spatial distributions of
hydrogen atoms and heavy elements.  While hydrogen gas is observed all
the way out to 500 kpc in projected distance with a mean covering
fraction of $\approx 60$\%, the same galaxies that display moderately
strong H\,I absorption exhibit few associated metal absorption lines
at $d > 200$ kpc.  The lack of heavy elements at large distances
persists through all ionization states included in the study,
suggesting that either there exists a chemical enrichment edge at
$d\approx 0.7\,R_h$ or gaseous clumps giving rise to the observed
absorption lines cannot survive at these large distances.

(2) The distinction between chemically-enriched and metal-poor gas is
further enhanced after accounting for a mass scaling of gaseous radius
in galactic halos.  We infer a halo mass $M_h$ and halo radius $R_h$
for each galaxy based on the known stellar mass $M_{\rm star}$, and
find that no ionic transitions are detected at $d\apg 0.7\,R_h$ to
sensitive upper limits of rest-frame absorption equivalent width
$W_r\approx 0.05$ \AA\ or better.  The observed absence of heavy elements at
$d\apg 0.7\,R_h$ applies to both low-mass dwarfs and high-mass
galaxies.  Considering all galaxies at $d>R_h$, we place a
strict {\it upper limit} for the covering fraction of heavy elements
of 3\% (at a 95\% confidence level) over the projected distance range
from $R_h$ to $9\,R_h$.

(3) To further improve upon the sensitivity in searching for weak
absorption features, we experiment with co-adding the absorption
spectra of individual galaxies in the rest frame.  Using stacked
spectra we are indeed able to improve the constraints for the
presence/absence of absorbing gas by factors of five to six. No ionic
transitions are found at $d > R_h$ to unprecedentedly sensitive limits
of $\approx 0.03$ \AA\ or better at the 95\% confidence level.

(4) Within the gaseous halo probed by our galaxy-QSO pair sample, we
observe differential covering fraction between low- and
high-ionization gas.  The mean covering fractions of both low- and
high-ionization absorbing gas remain high at $\mkap\apg 50$\% over the
distance range of $d=(0.25-0.6)\,R_h$.  At $d=(0.6-1)\,R_h$, the
covering fractions of C\,II and Si\,III decline to $\mkap\apl 14$\%,
while the covering fraction of \civ\ remains at $\mkap_{\rm
  C\,IV}=38$\%.  Combining previously observed high covering fraction
of low-ionization gas in inner halos ($d<0.2\,R_h$) with our findings
suggests that halo gas becomes progressively more ionized from
$d<0.2\,R_h$ to larger distances.

(5) Combining our study of outer gaseous halos of low- and high-mass
galaxies with the observations of the inner halos around massive
galaxies from the COS-Halos program establishes a consistent
$R_h$-normalized spatial absorption profile of the low-redshift CGM.
Both the absorption strengths and incidence (covering fraction) of
heavy elements decline steeply beyond $0.3\,R_h$ and become
vanishingly small beyond $0.7\,R_h$.  In contrast, massive starburst
galaxies at $z=2.2$ exhibit significantly stronger mean absorption
than galaxies at $z\sim 0$.  The differences are most pronounced in
low-ionization species traced by C\,II and Si\,II absorption features,
suggesting distinct ionization conditions between the CGM at low and
high redshifts.

(6) We caution a potential pseudo-evolution in the CGM radial
absorption profiles, when utilizing $R_h$-normalized projected
distances to account for the intrinsic size difference between
galaxies of different halo mass.  The commonly adopted spherical
overdensity calculation for halo mass imposes a pseudo-evolution for
static halos as the universe expands and the mean matter density
decreases with time.  Normalzing the observed CGM radial profile by
the scale radius $r_s$ or by the stellar mass $M_{\rm star}$ confirms
the previous finding that no significant changes can be found between
the CGM at $z\approx 2.2$ and $z\approx 0$.

\section*{Acknowledgments}

This work has benefited from valuable discussions with Benedikt
Diemer, Jean-Ren\'e Gauthier, Nick Gnedin, Sean Johnson, and Andrey
Kravtsov.  We thank Andrey Kravtsov for providing the $M_{\rm
  star}-M_h$ relation during the preparation of this paper, Michael
Blanton for referring us to the NASA-Sloan Atlas, and Alice Shapley
and Naveen Reddy for providing a representative stellar mass
distribution of the galaxy sample included in the study of Steidel et
al.\ (2010).  We also thank an annonymous referee for insightful
comments that helped improve the presentation of the paper.  This
research has made use of the NASA/IPAC Extragalactic Database (NED)
which is operated by the Jet Propulsion Laboratory, California
Institute of Technology, under contract with the National Aeronautics
and Space Administration.  Some of the data presented in this paper
were obtained from the Mikulski Archive for Space Telescopes
(MAST). STScI is operated by the Association of Universities for
Research in Astronomy, Inc., under NASA contract NAS5-26555. Support
for MAST for non-HST data is provided by the NASA Office of Space
Science via grant NNX13AC07G and by other grants and contracts.


\begin{center}
\begin{footnotesize}
\begin{table*}

 \centering

 \begin{minipage}{160mm}
   \caption{Summary of Galaxy Properties$^a$}

\begin{tabular}{|l|r|r|r|r|r|r|r|r|c|}
\hline
\hline
  \multicolumn{1}{|c|}{Galaxy Name} &
  \multicolumn{1}{c|}{RA(J2000)} &
  \multicolumn{1}{c|}{Dec(J2000)} &
  \multicolumn{1}{c|}{$z_{\rm gal}$} &
  \multicolumn{1}{c|}{$\log\frac{{\rm SFR}^b}{{\rm M}_\odot\,/\,{\rm yr}}$} &
  \multicolumn{1}{c|}{$\log\frac{M_{\rm star}}{{\rm M}_{\odot}}$} &
  \multicolumn{1}{c|}{$\log\frac{M_h^c}{M_{\odot}}$} &
  \multicolumn{1}{c|}{$R_h$(kpc)} &
  \multicolumn{1}{c|}{$M_{{\rm SDSS},r}^d$} &
   \multicolumn{1}{c|}{${\rm Flag}^e$} \\
\hline
  SDSS\,J000545.07$+$160853.3       & 00:05:45.07 &     $+$16:08:53.26  & 0.0372        & $-$0.76   & 9.3   & 11.2 & 137.0 & $-$18.8  & 0 \\
  SDSS\,J000548.11$+$161123.4       & 00:05:48.12 &     $+$16:11:23.40  & 0.1134        & $...$     & 10.5  & 11.9 & 217.1 & $-$21.2  & 1 \\
  2MASX\,J000556.17$+$160804.1      & 00:05:56.16 &     $+$16:08:04.17  & 0.0909        & $...$     & 10.7  & 12.0 & 241.5 & $-$21.5  & 1 \\
  SDSS\,J004208.22$-$102929.1       & 00:42:08.23 &     $-$10:29:29.10  & 0.0421        & $-$0.50   & 9.3   & 11.2 & 133.6 & $-$19.0  & 0 \\
  SDSS\,J004214.99$-$104414.9       & 00:42:14.99 &     $-$10:44:14.90  & 0.0360        & $-$0.88   & 9.5   & 11.3 & 144.9 & $-$19.6  & 0 \\
  SDSS\,J022617.97$+$001203.8       & 02:26:17.97 &     $+$00:12:03.85  & 0.0913        & $...$     & 9.8   & 11.4 & 153.8 & $-$19.7  & 1 \\
  2MASX\,J024248.74$-$082356.3      & 02:42:48.70 &     $-$08:23:56.65  & 0.0144        & $-$1.83   & 9.7   & 11.3 & 154.7 & $-$19.0  & 0 \\
  SDSS\,J025405.61$-$005236.6       & 02:54:05.64 &   $-$00:52:36.35  & 0.0032  & $-$2.33   & 7.1   & 10.0 & 55.8  & $-$13.9  & 0 \\
  SDSS\,J025906.86$+$004306.4       & 02:59:06.87 &   $+$00:43:06.34  & 0.0432  & $-$1.19   & 8.9   & 11.0 & 118.2 & $-$17.7  & 0 \\
  SDSS\,J025914.53$+$003359.6       & 02:59:14.53 &   $+$00:34:00.73  & 0.0092  & $-$2.44     & 7.2   & 10.1 & 60.6  & $-$15.5  & 0 \\
  SDSS\,J025938.43$+$003939.3       & 02:59:38.44 &   $+$00:39:39.24  & 0.0311  & $-$1.53     & 8.0   & 10.5 & 83.0  & $-$16.8  & 0 \\
  2MASX\,J025942.96$+$003907.8      & 02:59:42.94 &   $+$00:39:08.19  & 0.1342  & $...$     & 11.0  & 12.4 & 327.2 & $-$22.2  & 1 \\
\hline    
\multicolumn{10}{l}{$^a$ The full table in machine-readable form is available at http://lambda.uchicago.edu/public/local/galaxy\_table.dat.} \\
\multicolumn{10}{l}{$^b$ Star formation rate derived from GALEX Near UV band with $\lambda_{\rm eff}\approx$ 2267 \AA.}\\
\multicolumn{10}{l}{$^c$ The halo mass $M_h$ and halo radius $R_h$  are estimated following the prescription described in \S\ 4.3.}\\
\multicolumn{10}{l}{$^d$ Absolute $r$ band magnitude estimated from Sergic $+$ Exponential light profile fit (Bernadi et al.\ 2009). }\\
\multicolumn{10}{l}{$^e$ Galaxies with $M_{\rm star}$ from the NASA-Sloan Atlas are noted as ``0'', and galaxies with $M_{\rm star}$ inferred from Equation (1) are noted as ``1''.}
\label{galaxy_table}
\end{tabular}
\end{minipage}
\end{table*}
\end{footnotesize}
\end{center}

\begin{center}
\begin{footnotesize}
\begin{table*}
\centering
\begin{minipage}{160mm}
   \caption{Summary of Observations of QSOs$^a$}
\begin{tabular}{|l|r|r|r|r|r|r|r|}
\hline
\hline
  \multicolumn{1}{|c|}{QSO Name} &
  \multicolumn{1}{c|}{RA(J2000)} &
  \multicolumn{1}{c|}{Dec(J2000)} &
  \multicolumn{1}{c|}{$z_{\rm QSO}$} &
  \multicolumn{1}{c|}{PID} &
  \multicolumn{1}{c|}{$S/N_{\rm G130M}^b$} &
  \multicolumn{1}{c|}{$S/N_{\rm G160M}^b$} &
  \multicolumn{1}{c|}{$S/N_{\rm STIS}^c$} \\
\hline
  PG\,0003$+$158              & 00:05:59.24 & $+$16:09:49.01 & 0.4509   & 12038 & 16   & 13   & $...$\\
  SDSS\,J004222.29$-$103743.8 & 00:42:22.29 & $-$10:37:43.79 & 0.4240   & 11598 & 6    & 4    & $...$\\
  SDSS\,J022614.46$+$001529.7 & 02:26:14.44 & $+$00:15:30.04 & 0.6151   & 11598 & 8    & 5    & $...$\\
  SDSS\,J024250.85$-$075914.2 & 02:42:50.86 & $-$07:59:14.25 & 0.3777   & 12248 & 6    & 5    & $...$\\
  SDSS\,J025937.46$+$003736.3 & 02:59:37.47 & $+$00:37:36.38 & 0.5342   & 12248 & 6    & 5    & $...$\\
  PKS\,0312$-$770             & 03:11:55.25 & $-$76:51:50.85 & 0.2252   & 8651  & $...$ & $...$ & 5\\
  SDSS\,J040148.98$-$054056.5 & 04:01:48.98 & $-$05:40:56.58 & 0.5701   & 11598 & 6    & 5    & $...$\\
  PKS\,0405$-$123             & 04:07:48.43 & $-$12:11:36.66 & 0.5726   & 11541 & $...$ & $...$ & 48\\
  RX\,J0439.6$-$5311          & 04:39:38.72 & $-$53:11:31.40 & 0.2430   & 11520 & 11   & 7    & $...$\\
  HE\,0439$-$5254             & 04:40:12.02 & $-$52:48:17.70 & 1.0530   & 11520 & 11   & 8    & $...$\\

\hline
  \multicolumn{8}{l}{$^a$ The full table in machine-readable form is available at http://lambda.uchicago.edu/public/local/qso\_table.dat.} \\
  \multicolumn{8}{l}{$^b$Median $S/N$ of COS G130M grating in spectral range of $1150-1450$\AA \space and G160M grating in $1450-1750$\AA.} \\
  \multicolumn{8}{l}{$^c$Median $S/N$ of STIS spectra in spectral range of $1200-1700$\AA.} \\
\end{tabular}
\end{minipage}
\end{table*}
\end{footnotesize}
\end{center}


\begin{footnotesize}
\begin{table*}
\begin{center}
 \footnotesize
\centering
\begin{minipage}{160mm}
   \caption{Available Absorption Constraints of Different Galaxy Samples}
   \begin{tabular}{@{}lccccr@{}}
     \hline
     \hline
     \multicolumn{1}{c}{}& \# of Galaxies & $\langle z \rangle^a$  & $d$ Range  probed (kpc)  & \multicolumn{1}{c}{$\langle\,\log{M_{\rm star} / M_\odot}\,\rangle^b$}  & \multicolumn{1}{c}{Transitions$^c$} \\
     \hline
     This work 			& 195 	& 0.04 $\pm$ 0.04 	&	$32-500$  	&9.7  $\pm$ 1.1  & Ly$\alpha$, C\,II, C\,IV, Si\,II, Si\,III, Si\,IV  \\ 
     
     Werk et al.\ 2013 		& 44 		& 0.22 $\pm$ 0.05	&	$18-154$  	& 10.6 $\pm$ 0.5 & Ly$\alpha$, C\,II, Si\,II, Si\,III, Si\,IV  \\ 
     
     Steidel et al.\ 2010  	& 512 	& 2.2 $\pm$ 0.3 		&	$10-125(280)^d$ & 9.9 $\pm$ 0.5$^e$ & Ly$\alpha$, C\,II, C\,IV, Si\,II, Si\,IV  \\    
     \hline    
     \multicolumn{6}{l}{$^a$Median redshift and dispersion of the galaxy samples.}\\
     \multicolumn{6}{l}{$^b$Median stellar mass and dispersion of the galaxy samples.}\\
    \multicolumn{6}{l}{$^c$Including only transitions common to this work.}\\
    \multicolumn{6}{l}{$^d$Maximum impact parameter is 280 kpc for Ly$\alpha$ and 125 kpc otherwise. }\\
     \multicolumn{6}{l}{$^e$Median and dispersion of stellar mass for Steidel et al. 2010, which is a representative sample of galaxies in Reddy et al. 2012.}\\
   \label{line_table}
    
 \end{tabular}
\end{minipage}
\end{center}
\end{table*}
\end{footnotesize}

\begin{tiny}

\begin{sidewaystable*}

 \centering
 \vspace*{18cm}
   \hspace*{-1.5cm}
 \begin{minipage}{180mm}
   \begin{tabular}{@{}lrrccrrrrrr@{}}
   	\multicolumn{10}{l}{\textbf{Table 4. Summary of Individual Halo Absorption Properties.$^{a,b}$ }} \\
\hline
\hline
      \multicolumn{1}{c}{}& \multicolumn{1}{c}{} & \multicolumn{1}{c}{}  &  & & \multicolumn{1}{c}{H\,I}  & \multicolumn{1}{c}{Si\,III} &  \multicolumn{1}{c}{Si\,II} &  \multicolumn{1}{c}{Si\,IV} & \multicolumn{1}{c}{C\,II} & \multicolumn{1}{c}{\civ} \\
      \multicolumn{1}{c}{}& \multicolumn{1}{c}{$\theta$} & \multicolumn{1}{c}{$d$}  &  & & \multicolumn{1}{c}{$W_r(1215)$}  & \multicolumn{1}{c}{$W_r(1206)$} &  \multicolumn{1}{c}{$W_r(1260)$} &  \multicolumn{1}{c}{$W_r(1393)$} & \multicolumn{1}{c}{$W_r(1334)$} & \multicolumn{1}{c}{$W_r(1548)$} \\
               \multicolumn{1}{c}{Galaxy}        &     \multicolumn{1}{c}{($''$)}        & \multicolumn{1}{c}{(kpc)}      &         \multicolumn{1}{c}{$z_{\rm gal}$}               &             \multicolumn{1}{c}{$z_{\rm Ly\alpha}$}                  & \multicolumn{1}{c}{(m\AA)}  & \multicolumn{1}{c}{(m\AA)} &  \multicolumn{1}{c}{(m\AA)} & \multicolumn{1}{c}{(m\AA)} & \multicolumn{1}{c}{(m\AA)} & \multicolumn{1}{c}{(m\AA)} \\
\hline
	SDSS\,J112644.33$+$590926.0		&	393.2	&	32.4	&	0.0040	&	0.0040	&	993		$\pm$	15	&	290	$\pm$	11	&	$...$			&	166	$\pm$	12	&	106	$\pm$	12	&	586	$\pm$	12\\
	SDSS\,J134249.99$-$005329.0		&	29.1		&	39.3	&	0.0708	&	0.0717	&	660		$\pm$	16	&	78	$\pm$	6	&	69	$\pm$	8	&	$\leq$	16		&	140	$\pm$	14	&	309	$\pm$	24\\
	NGC\,3485						&	582.5	&	57.6	&	0.0048	&	0.0048	&	506		$\pm$	12	&		$\leq$	22	&	$\leq$	11		&	$\leq$	21		&	$\leq$	35		&	$\leq$	22\\
	SDSS\,J144520.23$+$341948.1		&	529		&	60.6	&	0.0056	&	0.0056	&	962		$\pm$	94	&		$\leq$	38	&	$\leq$	20		&	$\leq$	36		&	84	$\pm$	19	&	266	$\pm$	28\\
	2MASX\,J08091327$+$4618424		&	68.6		&	62.7	&	0.0466	&	0.0464	&	1026	$\pm$	10	&	54	$\pm$	6	&	$\leq$	11		&	69	$\pm$	7	&	$\leq$	21		&	$\leq$	27\\
	SDSS\,J122815.96$+$014944.1		&	1107.1	&	69.5	&	0.0030	&	0.0034	&	393		$\pm$	8	&		$\leq$	7	&	$\leq$	4		&	$\leq$	3		&	$\leq$	5		&	$\leq$	8\\
	SDSS\,J121413.94$+$140330.4		&	56.6		&	70.1	&	0.0644	&	0.0644	&	859		$\pm$	9	&		$\leq$	6	&	$\leq$	4		&	$\leq$	10		&	$\leq$	10		&	$\leq$	15\\
	2MASX\,J043936.88$-$530045.5		&	646.1	&	74.3	&	0.0056	&	0.0055	&	518		$\pm$	9	&	152	$\pm$	10	&	$...$			&	47	$\pm$	8	&	$\leq$	17		&	59	$\pm$	10\\
	SDSS\,J025914.53$+$003359.6		&	406		&	76.6	&	0.0092	&	0.0094	&	446		$\pm$	33	&		$...$		&	$\leq$	15		&	$\leq$	29		&	$\leq$	33		&	$\leq$	39\\
	SDSS\,J025938.43$+$003939.3		&	123.7	&	77.0	&	0.0311	&	0.0311	&	196		$\pm$	12	&		$\leq$	21	&	$\leq$	28		&	$\leq$	25		&	$\leq$	31		&	$\leq$	31\\
     \hline    
        \multicolumn{10}{l}{$^a$The full table in machine-readable form is available at http://lambda.uchicago.edu/public/local/absorber\_table.dat.} \\
      \multicolumn{10}{l}{$^b$Upper limits are quoted at 2 $\sigma$ level }\\ 
 
   \label{pair_table0}

 \end{tabular}
\end{minipage}

\end{sidewaystable*}
\end{tiny}

\setcounter{table}{4}


\begin{footnotesize}
\begin{table*}
 \footnotesize
 \centering
   \hspace*{1.5cm}
 \begin{minipage}{160mm}
   \caption{Absorption Equivalent Widths Observed in Stacked CGM Spectra in Different Projected Distance Range$^a$}
   \begin{tabular}{@{}rrccccccc@{}}
     
     \hline
     \hline
    \multicolumn{1}{c}{$d$} & \multicolumn{1}{c}{$\langle d \rangle ^b$} & & Ly$\alpha$  $\lambda$1215   &  Si\,III  $\lambda$1206  & Si\,II  $\lambda$1260  & Si\,IV  $\lambda$1393 & C\,II  $\lambda$1334 & C\,IV  $\lambda$1548 \\
     \multicolumn{1}{c}{(kpc)} & (kpc) & Number & (m\AA)  & (m\AA)   & (m\AA)  &  (m\AA) & (m\AA)   & (m\AA)\\
     \hline
	$32.4-138.1$	&  86.1	&	17	&	$	486	\pm	34	$	&	$	88	\pm	13	$	&	$	45	\pm	23	$	&	$	44	\pm	16	$ &	$	48	\pm	12	$	&	$	70	\pm	37	$	\\
	$156.2-307.1$ &  243.2	&	52	&	$	295	\pm	39	$	&	$		\le	7	$	&	$		\le	4	$	&	$		\le	13	$ &	$		\le	6	$	&	$	25	\pm	12	$	\\
	$311.0-442.4$ &  383.9	&	54	&	$	136	\pm	31	$	&	$		\le	3	$	&	$		\le	2	$	&	$		\le	4	$ &	$		\le	3	$	&	$		\le	8	$	\\
	$444.3-498.9$ &  480.7	&	53	&	$	116	\pm	23	$	&	$		\le	4	$	&	$		\le	6	$	&	$		\le	6	$ &	$		\le	5	$	&	$		\le	6	$	\\

     \hline    
 \multicolumn{9}{l}{$^a$ 1-$\sigma$ uncertainties and 2-$\sigma$ upper limits are determined from a bootstrap resampling calculation ($N_{\rm boot} = 350$)}\\ 
 \multicolumn{9}{l}{$^b$ Median impact parameter in each bin.}\\ 
   \label{pair_table1}
 \end{tabular}
\end{minipage}
\end{table*}
\end{footnotesize}

\begin{footnotesize}
\begin{table*}
 \footnotesize
 \centering
   \hspace*{1.5cm}
 \begin{minipage}{160mm}
   \caption{Absorption Equivalent Width in Stacked CGM Spectra in different $R_h$-Normalized Projected Distance Range$^a$}
   \begin{tabular}{@{}rrccccccc@{}}
     \hline
     \hline
     & &  & Ly$\alpha$  $\lambda$1215  &  Si\,III  $\lambda$1206  & Si\,II  $\lambda$1260  & Si\,IV  $\lambda$1393  & C\,II  $\lambda$1334 & C\,IV  $\lambda$1548 \\
     \multicolumn{1}{c}{$d/R_h$} &  \multicolumn{1}{c}{$\langle d/R_h \rangle $$^b$} & Number & (m\AA)  & (m\AA)   & (m\AA)  &  (m\AA) & (m\AA)   & (m\AA)\\
     \hline
$	0.25 - 0.54	$	&	0.43	&	8	&	$	550	\pm	70	$	&	$	93	\pm	32	$	&	$	67	\pm	25	$	&	$	204	\pm	94	$ &	$	102	\pm	56	$	&	$	164	\pm	86	$	\\
$	0.56 - 1.09	$	&	0.91	&	22	&	$	382	\pm	50	$	&	$	32	\pm	15	$	&	$		\le	4	$	&	$		\le	34	$ &	$		\le	6	$	&	$	63	\pm	26	$	\\
$	1.13 - 1.67	$	&	1.49	&	33	&	$	252	\pm	31	$	&	$		\le	4	$	&	$		\le	5	$	&	$		\le	20	$ &	$		\le	8	$	&	$		\le	31	$	\\
$	1.71 - 2.82	$	&	2.13	&	56	&	$	165	\pm	29	$	&	$		\le	3	$	&	$		\le	3	$	&	$		\le	9	$ &	$		\le	4	$	&	$		\le	4	$	\\
$	2.82 - 8.97	$	&	4.20	&	57	&	$	80	\pm	14	$	&	$		\le	10	$	&	$		\le	3	$	&	$		\le	3	$ &	$		\le	4	$	&	$		\le	6	$	\\
     \hline    
 \multicolumn{9}{l}{$^a$ 1-$\sigma$ uncertainties and 2-$\sigma$ upper limits are determined from a bootstrap resampling calculation ($N_{\rm boot} = 350$)}\\ 
 \multicolumn{9}{l}{$^b$ Median $R_h$-normalized projected distance in each bin.}\\ 
 \end{tabular}
\end{minipage}
\end{table*}
\end{footnotesize}

\begin{footnotesize}
\begin{table*}
 \footnotesize
 \centering
   \hspace*{1.5cm}
 \begin{minipage}{160mm}
   \caption{Mean Gas Covering Fraction, \mkap, of Different Species$^a$}
   \begin{tabular}{@{}cccccccc@{}}
     
     \hline
     \hline
$d/R_h$  & \multicolumn{1}{c}{$\langle d/R_h \rangle$$^b$} & Ly$\alpha$ $\lambda$1215 & Si\,III $\lambda$1206  & Si\,II $\lambda$1260  & Si\,IV $\lambda$1393 & C\,II $\lambda$1334 & C\,IV $\lambda$1548  \\
     \hline
$       0.25 - 0.54     $ &  0.43       & $1.00_{-0.18}                 $   & $0.63^{+0.13}_{-0.18}$    & $0.43^{+0.18}_{-0.15}$                & $0.50^{+0.18}_{-0.18}$        & $0.63^{+0.13}_{-0.18}$        & $0.60^{+0.16}_{-0.21}$  \\
$       0.56 - 1.09     $ &  0.91       & $0.96^{+0.01}_{-0.08} $   & $0.14^{+0.11}_{-0.05}$    & $\leq 0.12$                   & $0.05^{+0.09}_{-0.02}$  & $0.06^{+0.11}_{-0.02}$      & $0.38^{+0.13}_{-0.10}$  \\
$       1.13 - 1.67     $ &  1.49       & $0.94^{+0.02}_{-0.07} $   & $\leq 0.09$                       & $\leq 0.08$                   & $\leq 0.11$           & $\leq  0.12$                  & $\leq   0.12$                         \\
$       1.71 - 2.82     $ &  2.13       & $0.71^{+0.05}_{-0.07} $   & $\leq 0.07$                       & $\leq 0.06$                   & $\leq 0.06$           & $\leq  0.07$                  & $\leq   0.06$                         \\
$       2.82 - 8.97     $ &  4.20       & $0.59^{+0.07}_{-0.08} $   & $\leq 0.09$                       & $\leq 0.06$                   & $\leq 0.07$           & $\leq  0.06$                  & $\leq   0.06$                         \\
     \hline
 \multicolumn{8}{l}{$^a$ Based on the fraction of absorbers with rest-frame absorption equivalent width greater than $W_0 = 0.05$ \AA.}\\ 
 \multicolumn{8}{l}{$^b$ Median $R_h$-normalized projected distance in each bin.}\\ 
   \label{kappa_table}
 \end{tabular}
\end{minipage}
\end{table*}
\end{footnotesize}

\begin{footnotesize}
\begin{table*}
 \footnotesize
 \centering
   \hspace*{1.5cm}
 \begin{minipage}{160mm}
   \caption{Absorption Equivalent Width in Stacked Spectra of \lya\ Absorbers at $d>0.7\,R_h$$^a$}
   \begin{tabular}{@{}rrcccccccc@{}}
     \hline
     \hline
     & &  & Ly$\alpha$ 1215  &  N\,V 1238 & Si\,III 1206  & Si\,II 1260  & Si\,IV 1393  & C\,II 1334 & C\,IV 1548 \\
     \multicolumn{1}{c}{$d/R_h$} &  \multicolumn{1}{c}{$\langle d/R_h \rangle $$^b$} & Number & (m\AA)  & (m\AA)   & (m\AA)  &  (m\AA) & (m\AA)   & (m\AA)& (m\AA)\\
     \hline
$	0.96 - 1.63	$	&	1.26		&	32	&	$	308	\pm	28	$	& $\le 15	$ &	$	        \le     	5	$	&	$		\le	5	$	&	$	        \le  14  	$ &	$		\le	6	$	&	$	        \le   33	$	\\
$	1.64 - 2.59	$	&	1.85		&	31	&	$	281	\pm	34	$	& $\le 8	$  & 	$		\le	6	$	&	$		\le	8	$	&	$		\le 15	$ &	$		\le	15	$	&	$		\le  17	$	\\
$	2.59 - 8.97	$	&	4.06		&	30	&	$	131	\pm	19	$	& $\le 7	$ &	$		\le	9	$	&	$		\le	9	$	&	$		\le 8		$ &	$		\le	4	$	&	$		\le  6		$	\\
     \hline    
 \multicolumn{9}{l}{$^a$ 1-$\sigma$ uncertainties and 2-$\sigma$ upper limits are determined from a bootstrap resampling calculation ($N_{\rm boot} = 350$)}\\ 
 \multicolumn{9}{l}{$^b$ Median $R_h$-normalized projected distance in each bin.}\\ 
   \label{pair_table2}
 \end{tabular}
\end{minipage}
\end{table*}
\end{footnotesize}



\clearpage


\label{lastpage}

\end{document}